\documentclass{aastex62}

\usepackage{color}

\shorttitle{RMHD simulations of AGN}
\shortauthors{Igarashi et al.}
\begin{document}

%\title{Radiation Magnetohydrodynamic Simulations of Sub-Eddington Accretion Disks: Origin of Soft X-ray Excess of Changing Look AGN} 
\title{Radiation Magnetohydrodynamic Simulations of Sub-Eddington Accretion Flows in AGN: Origin of Soft X-ray Excess and Rapid Time Variabilities} 

\correspondingauthor{Taichi Igarashi}
\email{igarashi.taichi@chiba-u.jp}
\author{Taichi Igarashi}

\affiliation{Department of Physics, Graduate School of Science, Chiba University\\
1-33 Yayoi-Cho, Inage-Ku\\
Chiba 263-8522, Japan}

%\author[0000-0003-2349-9003]{Yoshiaki Kato}
\author{Yoshiaki Kato}
\affiliation{RIKEN \\
2-1 Hirosawa, Wako\\
Saitama, 351-0198, Japan}

\author{Hiroyuki R. Takahashi}
\affiliation{Faculty of Arts and Sciences, Department  of Natural Sciences, Komazawa University\\
1-23-1 Komazawa, Setagaya \\
Tokyo, 154-8525, Japan}

\author{Ken Ohsuga}
\affiliation{Center for Computational Sciences, University of Tsukuba\\
1-1-1 Ten-nodai\\
Tsukuba 305-8577, Japan}

\author{Yosuke Matsumoto}
\affiliation{Department of Physics, Graduate School of Science, Chiba University\\
1-33 Yayoi-Cho, Inage-Ku\\
Chiba 263-8522, Japan}

\author{Ryoji Matsumoto}
\affiliation{Department of Physics, Graduate School of Science, Chiba University\\
1-33 Yayoi-Cho, Inage-Ku\\
Chiba 263-8522, Japan}



\begin{abstract}
We investigate the origin of the soft X-ray excess component in Seyfert galaxies observed when their luminosity exceeds 0.1\% of the Eddington luminosity ($L_{\mathrm{Edd}}$).
%%X-ray observations indicate that soft X-ray excess component appears in Seyfert galaxies when their bolometric luminosity exceeds 0.1\% of the Eddington luminosity.
%
%%We present the results of global three-dimensional radiation magnetohydrodynamic (RMHD) simulations of accretion flows around a $10^7M_{\odot}$ black hole. 
%
The evolution of a dense blob in radiatively inefficient accretion flow (RIAF) is simulated by applying a radiation magnetohydrodynamic code, CANS+R. 
%
%General relativistic effects are simulated by using Pseudo-Newtonian potential.
%Pseudo-Newtonian potential is used to describe the gravitational fields around a Schwarzschild black hole. 
%%
When the accretion rate onto a $10^7M_{\odot}$ black hole exceeds 10\% of the Eddington accretion rate {($\dot M_{\rm Edd}=L_{\rm Edd}/c^2$, where $c$ is the speed of light)}, the dense blob shrinks vertically because of radiative cooling and forms a Thomson thick, relatively cool ($\sim10^{7-8}$ K) region. 
%{The cool region is the origin of soft X-ray emission observed in Seyfert galaxies.}
The cool region coexists with the optically thin, hot ($T\sim10^{11}~\mathrm{K}$) RIAF near the black hole. The cool disk is responsible for the soft X-ray emission, while hard X-rays are emitted from the hot inner accretion flow.
%
%The soft X-ray emitting region coexists with  the optically thin, hot ($T \sim 10^{11}~\mathrm{K}$), radiatively inefficient accretion flow (RIAF) near the black hole.
%%
Such a hybrid structure of hot and cool accretion flows is consistent with the observations of both hard and soft X-ray emissions from `changing-look' active galactic nuclei (CLAGN).
%
%Furthermore, we find that quasi-periodic oscillations (QPOs) with the timescale of days are excited {in the soft X-ray emitting region}.
Furthermore, we find that quasi-periodic oscillations (QPOs) are excited in the soft X-ray emitting region.
These oscillations can be the origin of rapid  X-ray time variabilities observed in CLAGN.

\end{abstract}


%% cross-references to sections, Equations, tables, and figures.

%%%%
\section{Introduction} \label{sec:intro}

{ Active galactic nuclei (AGN) are categorized into low-luminosity AGN such as M87 and luminous AGN such as Seyfert galaxies. In low luminosity AGN, their luminosity $L$ is less than 0.1\% of the Eddington luminosity $L_{\rm Edd}$. 
The broadband spectrum of electromagnetic radiation from low-luminosity AGN can be explained by radiatively inefficient accretion flows (RIAFs) onto a supermassive black hole \citep[e.g.,][]{narayan+1995}.
RIAFs are optically thin, hot ($T \sim 10^{11}$K), advection-dominated accretion flows in which the radiation energy is much smaller than the thermal energy. On the other hand, the optical-UV thermal radiation observed in luminous AGN is emitted by optically thick,  cold ($T \sim 10^{4-5}$K) disks. 
Such disks can be modeled by standard accretion disks \citep{shakura+sunyaev1973}} driven by  sub-Eddington accretion flows.  
A key parameter characterizing the activity of AGN is the ratio of their accretion rate $\dot M$ to the Eddington accretion rate $\dot M_{\rm Edd}$ defined by $\dot M_{\rm Edd}=L_{\rm Edd}/c^2$, where $c$ is the speed of light.

\citet{abramowicz+1995} showed that RIAFs can exist only when the accretion rate is less than 1-10\% of the Eddington accretion rate. 
When the accretion rate of a RIAF exceeds this upper limit, radiative cooling dominates the viscous heating so that the RIAF cools and transitions to an optically thick, cold disk. 
Such transitions are observed in stellar-mass black holes as hard-to-soft state transitions \citep[e.g.,][]{fender+2004}. 
In AGN, it is expected that low-luminosity AGN evolve toward luminouse AGN when the accretion rate exceeds the upper limit for RIAF. 
During this transition, accretion flow stays in an intermediate state between a RIAF and standard disks. 
The time evolution of such sub-Eddington accretion flows onto a supermassive black hole can be studied by radiation magnetohydrodynamic (RMHD) simulations. 

Global three-dimensional magnetohydrodynamic (MHD) simulations of nonradiative accretion flows \citep[e.g.,][]{hawley2000} showed that the mass accretion is driven by Maxwell stress enhanced by magnetic turbulence driven by magnetorotational instability (MRI).  
\citet{machida+2000} showed that magnetic fields enhanced by MRI buoyantly escape from the disk and form a magnetically active disk corona. 
Global two-dimensional RMHD simulations of black hole accretion flows have been initiated by \citet{ohsuga+2009} and \citet{ohsuga+mineshige2011}. 
Three-dimensional RMHD simulations have been performed for accretion flows with an accretion rate exceeding the Eddington accretion rate \citep[e.g.,][]{takahashi+2016,sadowski+narayan2015,jiang+2019super}. 
{\cite{jiang+2019sub} presented the results of RMHD simulations for sub-Eddington accretion flows, but the accretion rate in their simulations is close to $\sim 1\dot M_{\rm Edd}$.
Numerical simulations for lower accretion rates but exceeding the upper limit for RIAFs are challenging because we need to resolve a cool disk.
%
%%%
%%
\begin{figure*}
\plotone{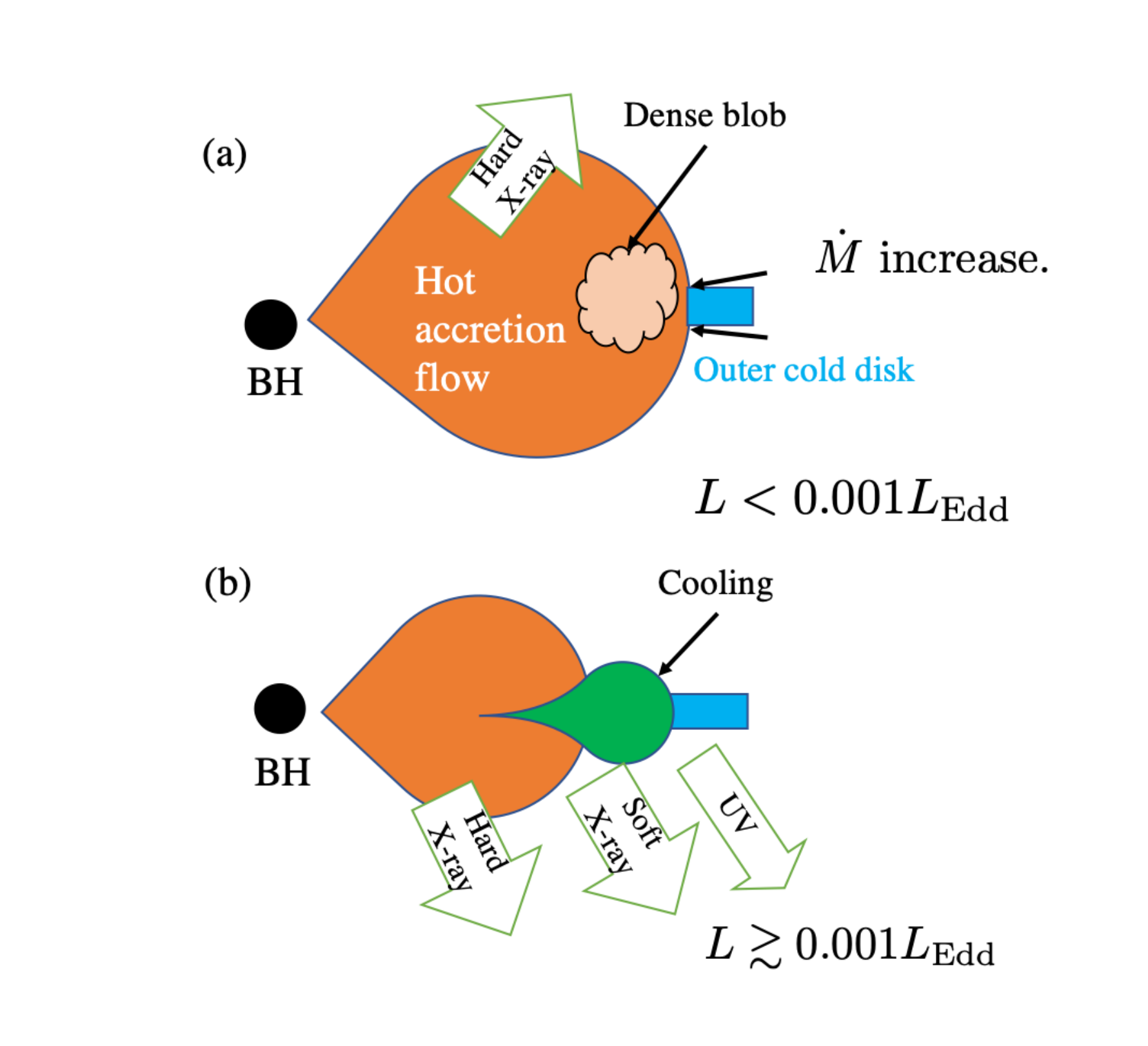}
\caption{A schematic picture of the central engine of AGN during the hard-to-soft transition. Orange, green, and blue areas show hot RIAF, soft X-ray emitting region and outer UV emitting standard disks, respectively. (a) A dense blob is formed by an increase of the mass accretion rate, (b) when the luminosity exceeds $0.001L_{\mathrm{Edd}}$, a soft X-ray emitting region is formed by radiative cooling.}
\label{fig:schematic}
\end{figure*}
\citet{machida+2008} carried out global three-dimensional MHD simulations of the nonradiative accretion flow infalling from the outer cool disk. 
They showed that when the cool matter infalls by losing angular momentum, an inner torus (or dense blob) is formed. 
Furthermore, they showed that the torus is deformed into a nonaxisymmetric shape due to the growth of the nonaxisymmetric instability \citep[e.g.,][]{papaloizou+pringle1984,drury1985}. 
They also showed that quasi-periodic oscillations (QPOs) are excited in the inner torus. 
Figure \ref{fig:schematic}(a) schematically shows a dense blob formed in the hot, low-density accretion flow. 
As the accretion rate from the outer cold disk increases, the blob density increases. 
When the density of the blob exceeds the upper limit for RIAF, the dense blob is subject to the cooling instability. 
Figure \ref{fig:schematic}(b) schematically shows an intermediate state between low-luminosity AGN and luminouse AGN \textbf{\citep[e.g.,][]{noda+2014,sniegowska+2020,mahmoud+2020}. }
In the region where the cool blob is formed by cooling instability, soft X-rays are emitted (the green region in Figure~\ref{fig:schematic}(b)).
%The radially stratified structure shown in Figure~\ref{fig;schematic}(b) is similar to the interface between the hot accretion flow and the outer cold, standard accretion disk in black hole binaries \citep[e.g.,][]{manmoto+2000}.
%It has been suggested that the time variabilities in the intermediate region can produce low frequency time variabilities in black hole binaries \citep[e.g.,][]{kato1989,mahmoud+done2018}.
%\textbf{In the case of stellar mass black holes, when the accretion rate from the outer cold disk increases, cooling instability will grow in the dense blob and forms a cool, magnetically sapported disk.
%On the other hand, in the case of AGN accretion disks, a soft X-ray emitting cool region (the green region in Figure \ref{fig:schematic}(b)).}
%%

%%
\textbf{
The radially stratified structure shown in Figure~\ref{fig:schematic}(b) is similar to the interface between the hot accretion flow and outer standard disk \citep[e.g.,][]{manmoto+2000}.
It has been suggested that such a region can produce a low-frequency oscillation in black hole binaries \citep[e.g.,][]{mahmoud+done2018}.
}
For stellar-mass black holes, \citet{machida+2006} presented the results of global three-dimensional MHD simulations of the hard-to-soft state transition.
\textbf{By considering optically thin radiative cooling, they showed that when the surface density of the disk exceeds the upper limit for RIAF, the disk shrinks in the vertical direction by radiative cooling and forms a region with temperature $10^8$ K outside the hot RIAF.}
They found that the azimuthal magnetic field is enhanced when the azimuthal component of the magnetic field is symmetric with respect to the equatorial plane of the disk.
Because this azimuthal magnetic field supports the disk, the disk stays in the optically thin state.
This intermediate state corresponds to the luminous hard state observed during the hard-to-soft state transition in stellar-mass black hole candidates.
\textbf{In addition, \citet{oda+2009,oda+2012} showed that when the azimuthal magnetic fields are considered, steady solutions of luminous hard-state disk exist even when the accretion rate exceeds the upper limit of RIAF.}
\citet{machida+2006} dismissed investigating further evolution because numerical instabilities grow in the magnetically dominated region.
More importantly, they could not treat the optically thick region that is formed in the stage later than their simulations because they assumed optically thin radiative cooling.

Some AGN show transitions between type 1 with broad emission lines and type 2 without broad {emission lines}.
Such transitions have been observed in Seyfert galaxies such as NGC2617 \citep[e.g.,][]{shappee+2014}, Mrk590 \citep[e.g.,][]{denney+2014}, Mrk1018 \citep{husemann+2016,lamassa+2017,noda+done2018}, and NGC1566 \citep{parker+2019,oknyansky+2019}.
Some quasars also show transitions between type 1 and type 2 \citep[e.g.,][]{lamassa2015,macleod+2016}, and these are called changing-look AGN (CLAGN).
The transitions occur in the timescale of years, disfavoring the model of an external origin, such as occultation by a dust torus.
CLAGN show soft X-ray time variations preceding the UV-optical intensity variations \citep[][]{shappee+2014,oknyansky+2017,noda+done2018}.
They also indicate that the primary source of the time variation is the soft X-ray emission from the region near the supermassive black hole.
Previous studies have shown that the X-ray spectra of Seyfert galaxies become softer as their luminosity increases \citep[e.g.,][]{arnaud+1985,turner+punds1989}.
Through a method of variability-assisted broadband spectroscopy, \citet{noda+2014} found that a highly variable soft X-ray excess component appears in NGC3227 when the source luminosity exceeds 0.1\% of the Eddington luminosity.
In contrast, \citet{noda+done2018} showed that when the bolometric luminosity in Mrk1018 decreases below 1\% of the Eddington luminosity, the soft X-ray excess disappears.
%

%%
%The appearance/disappearance of the soft X-ray excess component can be explained theoretically by considering thermal equilibrium curves of steady black hole accretion flows \citep{abramowicz+1995}.
%%
%The hard state without soft X-ray excess can be explained by radiatively inefficient accretion flows (RIAF).
%%
%RIAF are optically thin, hot, and advection-dominated flows \citep[e.g.,,][]{narayan+yi1995}.
%%
%There exists a gap between RIAFs and an optically thick disk.
%%
%Here, we should note that the soft X-ray-emitting region around a supermassive black hole is different from standard accretion disks \citep{shakura+sunyaev1973} in which the disk temperature is around $10^{5}$~K.
%%
%The soft X-ray excess region can rather be regarded as the intermediate state between RIAFs and standard disks.
%%

In this paper, we present the results of the RMHD simulation of the accretion flow when an overdense region appeared in RIAF, although the formation of such an overdense region in a realistic situation is left as an open question.
We consider accretion flows with accretion rate $\sim 0.1\dot M_{\rm Edd}$.
This mass accretion rate is an order of magnitude smaller than that in RMHD simulations reported by \citet{jiang+2019sub}.
We report the time-evolution of the structure and variability of the overdense region.

%%%%
\section{Method} \label{sec_method}

\subsection{Basic Equations}
We solve the RMHD Equations, which consist of resistive-magnetohydrodynamic Equations coupled with the 0th and 1st moments of radiation transfer Equations in cylindrical coordinates $(r,\varphi,z)$.
General relativistic effects are considered using the pseudo-Newtonian potential \citep{paczynsky+wiita1980} $\phi_{\mathrm{PN}} = -GM_{\mathrm{BH}}/(R-r_\mathrm{s})$, where $M_{\mathrm{BH}}=10^7M_{\odot}$ is the black hole mass, $R (=\sqrt{r^2+z^2})$ is the distance from the black hole, $r_\mathrm{s} (= 2GM_{\mathrm{BH}}/c^2)$ is the Schwarzschild radius, and $G$ is the gravitational constant.
\begin{table}
\begin{center}
\caption{Units in this paper}
\begin{tabular}{c|c}
\tableline
Units & Values \\ \tableline
Length & $r_{\mathrm s}=3\times10^{12}\ \mathrm{cm}$ \\
Velocity & $c=3.0\times10^{10}$ cm/s \\
Time & $t_0=r_{\mathrm s}/c=100\ \mathrm s$ \\
Luminosity & $L_{\mathrm{Edd}}=4\pi cGM_{\mathrm{BH}}/\kappa_{\mathrm{es}}=1.25\times10^{45}$ erg/s \\
Eddington accretion rate & $\dot{M}_{\mathrm{Edd}}=L_{\mathrm{Edd}}/c^2=1.4\times10^{24}$ g/s \\
Density & $\rho_0=2.0\times10^{-11}\ \mathrm{g/cm^3}$ \\
Magnetic Field & $B_0=\sqrt{4\pi\rho_0c^2}=4.9\times10^{5}$ G \\
\tableline
\end{tabular}
\end{center}
\label{tab:norm}
\end{table}
%%

%
%The black hole mass $M_{\mathrm{BH}}$ is fixed to $10^7M_\odot$, where $M_\odot$ is the solar mass.
%
The basic Equations are nondimensionalized by using normalizations shown in Table~\ref{tab:norm} and expressed as follows:
\begin{equation}
\frac{\partial\rho}{\partial t} + \mathbf{\nabla}\cdot(\rho \mathbf{v}) = 0,
\label{eqn:continuity}
\end{equation}
\begin{equation}
\frac{\partial\rho \mathbf{v}}{\partial t} + \mathbf{\nabla}\cdot(\rho \mathbf{vv} + p_{\mathrm t}\mathbf{I}- \mathbf{BB} ) =-\rho\mathbf{\nabla}\phi_\mathrm{PN} -\mathbf{S},
\label{eqn:motion}
\end{equation}
\begin{equation}
\frac{\partial {E_{\mathrm t}}}{\partial t} + \mathbf{\nabla}\cdot[(E_{\mathrm t}+p_{\mathrm t} )\mathbf{v} - \mathbf{B(v\cdot B)}] = -\mathbf{\nabla}\cdot(\eta\mathbf{j}\times\mathbf{B})-\rho\mathbf{v\cdot\nabla}\phi_{\rm{PN}} - c S_\mathrm{0},
\label{eqn:energy}
\end{equation}
\begin{equation}
\frac{\partial \mathbf{B}}{\partial t} + \mathbf{\nabla\cdot}(\mathbf{vB-Bv}+\psi\mathbf{I}) = -\mathbf{\nabla}\times(\eta\mathbf{j}),
\label{eqn:induction}
\end{equation}
\begin{equation}
\frac{\partial \psi}{\partial t} + c^2_{\mathrm h}\mathbf{\nabla\cdot\mathbf{B}} = -\frac{c^2_{\mathrm h}}{c^2_{\mathrm p}}\psi,
\label{eqn:divergencefree}
\end{equation}
where $\rho$, $\mathbf{v}$, $\mathbf{B}$, and $\mathbf{j} = \mathbf{\nabla} \times\mathbf{B}$ are the mass density, velocity, magnetic field, and current density, respectively. 
In addition, $p_{\mathrm t}=p_\mathrm{gas} + B^2/2$ is the total pressure, and $E_{\mathrm t} = \rho v^2 /2+ p_\mathrm{gas}/(\gamma - 1) + B^2/2$ is the total energy, where $\gamma=5/3$ is the specific heat ratio.
%
%We apply the so-called anomalous resistivity in which the resistivity $\eta$ becomes large when the electric current $\mathbf{j} = \mathbf{\nabla} \times\mathbf{B}$ exceeds the critical value \citep[e.g.,,][]{1994ApJ...436L.197Y}.
%
We apply the so-called anomalous resistivity:
\begin{equation}
	\eta =
	\left\{
	\begin{array}{l}
	\eta_0\min\left[1,\left(v_{\mathrm d}/v_{\mathrm c}-1\right)^2\right],\ v_{\mathrm d}\geq v_{\mathrm c} \\
	0,\ v_{\mathrm d} \leq v_{\mathrm c}
	\end{array}
	\right.
\end{equation}
where $\eta_0 = 0.01cr_{\mathrm s}$, $v_{\mathrm c}=0.9c$, and \textbf{$v_{\mathrm d}=jm_{\mathrm p}/(e\rho)$ are the upper limit of the resistivity, critical velocity, and drift velocity, respectively. 
Here $m_{\mathrm p}$ is the proton mass and $e$ is the electron charge.} 
The resistivity $\eta$ becomes large when the drift velocity $v_{\mathrm d}$ exceeds the critical velocity $v_{\mathrm c}$ \citep[e.g.,][]{1994ApJ...436L.197Y}.
In Equations~(\ref{eqn:induction}) and (\ref{eqn:divergencefree}), $\psi$ is introduced so that the divergence-free magnetic field is maintained within minimal errors during time integration where $c_\mathrm{h}$ and $c_\mathrm{p}$ are constants \citep[see][for more details]{matsumoto+2019}.

In Equations~(\ref{eqn:motion}) and (\ref{eqn:energy}), {$\mathbf{S}$ and $S_\mathrm{0}$ are the radiation momentum and the radiation energy source terms, respectively, and are derived by solving the frequency-integrated 0th and 1st moments of the radiation transfer Equations expressed in the following forms \citep[see][]{lowrie+1999,takahashi+2013,takahashi+ohsuga2013,kobayashi+2018}:
\begin{equation}
\frac{\partial E_\mathrm{r}}{\partial t} + \nabla\cdot \mathbf{F}_\mathrm{r} = cS_\mathrm{0},
\end{equation}
\begin{equation}
\frac{1}{c^2}\frac{\partial\mathbf{F}_\mathrm{r}}{\partial t} + \mathbf{\nabla}\cdot\mathbf{P}_\textrm{r} = \mathbf{S},
\end{equation}
where
\begin{equation}
S_\mathrm{0} = \rho\kappa_\mathrm{ff}(a_\mathrm{r} T^{4} - E_\mathrm{r}) + \rho(\kappa_\mathrm{ff}-\kappa_\mathrm{es})\frac{\mathbf{v}}{c}\cdot[\mathbf{F}_\mathrm{r} - (\mathbf{v}E_\mathrm{r} + \mathbf{v}\cdot\mathbf{P}_\mathrm{r})],
\end{equation}
\begin{equation}
\mathbf{S} = \rho\kappa_\mathrm{ff}\frac{\mathbf{v}}{c}(a_\mathrm{r}T^4 - E_\mathrm{r}) - \rho(\kappa_\mathrm{ff}+\kappa_\mathrm{es})\frac{1}{c}[\mathbf{F}_\mathrm{r} - (\mathbf{v}E_\mathrm{r} + \mathbf{v\cdot P}_\mathrm{r})].
\end{equation}
In Equations (9) and (10), $\kappa_\mathrm{ff} = 1.7\times10^{-25}m_\mathrm{p}^{-2}\rho T^{-7/2}~\mathrm{cm}^2/\mathrm{g}$ is the free-free absorption opacity where $\kappa_\mathrm{es} = 0.4~\mathrm{cm}^2/\mathrm{g}$ is the electron scattering opacity.
The gas temperature, $T$, is related to the gas pressure and density by $p_{\rm gas} = \rho k_{\mathrm B}T/(\mu m_{\rm p})$, where $k_{\mathrm B}$, and $\mu =0.5$ are the Boltzmann constant and mean molecular weight, respectively.
{In this paper, we mainly study the region with temperature higher than $10^7$~K. 
For more realistic simulations of AGN accretion flows including the lower temperature region, we have to consider the opacity of metals \citep[e.g.,][]{jiang+2019sub,jiang+2019super}.}
%
%{ These Equations} are solved with { M1} closure relation \citep[][and reference therein]{gonzalez+2007, takahashi+2016} .
%
%

%% Numerical procedure
The MHD part of the RMHD Equations is solved by CANS+ \citep{matsumoto+2019}, which adopts the HLLD  approximate Riemann solver \citep{2005JCoPh.208..315M}, fifth-order monotonicity-preserving interpolation scheme \citep{sureesh+huynh1997}, and hyperbolic divergence cleaning method for solving the induction Equation \citep{dedner+2002}. 
The simulation code for the radiation part is the same as \citet{kobayashi+2018}, and the moment Equations of the radiative transfer are solved with the M1-closure relation \citep[][and references therein]{gonzalez+2007, takahashi+2016}.
Our RMHD code is referred to as CANS+R hereafter.
%
%%\UTF{00DF}\UTF{00D4}\UTF{00C7}\UTF{00E1}%
\subsection{Initial condition and boundary conditions}
{We simulate the AGN accretion flows shown in Figure~\ref{fig:schematic}. 
It is desirable to include the outer cold disk and study the evolution of the inner region when the accretion rate from the outer cold disk increases.
In CLAGN, however, the time scale of the increase of the mass accretion rate that triggers the state transition of the inner region is several years.
This time scale is much longer than the time scale for which we can carry out global three-dimensional simulations including the region near the black hole.
Thus, we do not include the outer cold disk in the simulation region.
Instead, we start from a nonaccreting disk with temperature $T\sim10^8$~K and simulate the evolution of the disk without including the radiative cooling until the disk is heated by accretion driven by the angular momentum transport by MRI.
It should be noted that we do not simulate the origin of the increase of the accretion rate from the outer region, inducing the state transition in the inner region.
Instead, we study the evolution of the density enhancement produced by the accretion from the cold disk.}
\begin{figure*}
\plotone{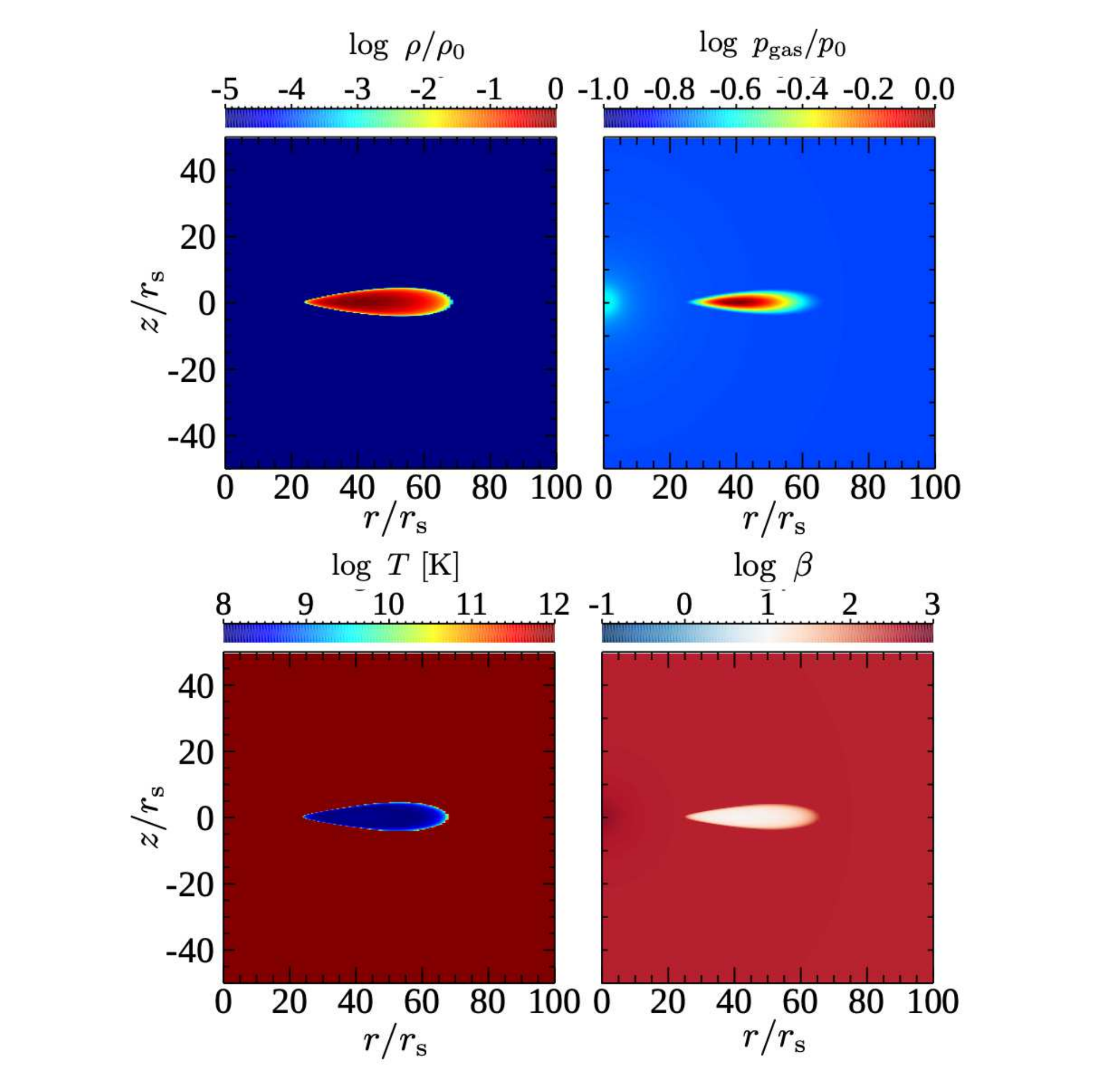}
\caption{Distribution of gas density (upper left), gas pressure (upper right),  gas temperature (lower left), and plasma $\beta$ (lower right) in the poloidal plane at $t=0$.}
\label{fig:init}
\end{figure*}
\begin{table}
\begin{center}
\caption{Parameters adopted in this paper.}
\begin{tabular}{c|c}
\tableline
Parameters & Values \\ \tableline
Radial dependence of angular momentum & $a=0.46$ \\
Radius at the initial density maximum & $r_0=40r_{\mathrm s}$ \\
Initial plasma $\beta$ around the density maximum & $\beta = 10$ \\
%$E_{\mathrm{th0}}=c_{\mathrm s0}^2/\gamma |\psi_0|$& \\
Sound speed of the disk & $c_{\mathrm s0}=5.6\times10^{-3}c$ \\
%$E_{\mathrm{thc}}=c_{\mathrm sc}^2/\gamma |\psi_0|$& \\
Sound speed of the corona & $c_{\mathrm sc}=0.9c$ \\
Gas pressure at the initial density maximum & $p_0=\rho_0c_{\mathrm s0}^2=6.0\times10^5\ \mathrm{g/cm^2/s}$ \\
%Initial density at $(r,z) = (40r_{\mathrm s},0)$ &$\rho_0 = 2\times10^{-11}\ \mathrm{g/cm^3}$ \\
Ratio of the density of the corona to the disk & $\rho_{\mathrm c0}/\rho_0=7\times10^{-6}$ \\
Black hole mass &$M_{\mathrm{BH}}=10^7M_{\odot}$ \\
\tableline
\end{tabular}
\end{center}
\label{tab:param}
\end{table}
For the initial condition, we set an equilibrium model of a rotating disk assuming a polytropic equation of state, $p=K\rho^{1+1/n}$ with $n=3$, and power-law specific angular momentum distribution, $l=l_0(r/r_0)^a$ with $l_0=\sqrt{GMr_0^3/(r_0-r_{\mathrm s})^2}$, where $r_0$ is the radius of the density maximum of the disk. 
The initial density and gas pressure of the disk are given by 
\begin{equation}
\rho_{\mathrm d} = \rho_0\left(1-\frac{\gamma}{c_{\mathrm s0}^2}\frac{\tilde{\phi}-\tilde{\phi_0}}{n+1}\right)^n
\end{equation}
and
\begin{equation}
%p_{\mathrm d} = \rho_0\frac{c_{\mathrm s0}^2}{\gamma}\left( \frac{\rho_{\mathrm d}}{\rho_0} \right)^{1+\frac{1}{n}}
p_{\mathrm d} =p_0\left( \frac{\rho_{\mathrm d}}{\rho_0} \right)^{1+\frac{1}{n}}
\end{equation}
respectively, where $\rho_0$, $c_{\mathrm s0}$ and $p_0=\rho_0c_{\mathrm s0}^2/\gamma$ are the initial density, sound speed and gas pressure of the disk at $(r,z)=(r_0,0)$, respectively.
The potential $\tilde{\phi}$ is the effective potential calculated by 
\begin{equation}
\tilde{\phi} = \phi+\frac{1}{2(1-a)}\left(\frac{l}{r}\right)^2.
\end{equation}
In Equation (11), $\tilde\phi_0 = \tilde\phi(r_0,0)$.
Outside the disk, we assumed a spherical, isothermal and static corona with temperature $T\sim10^{12}~\mathrm{K}$.
The density and pressure distribution of the corona are given by
\begin{equation}
\rho_{\mathrm c}=\rho_{\mathrm c0}\exp{\left(-\frac{\phi-\phi_{\mathrm 0}}{c_{\mathrm{sc}}^2/\gamma}\right)}
\end{equation}
and
\begin{equation}
p_{\mathrm c} = \rho_{\mathrm c}c_{\mathrm{sc}}^2/\gamma=p_{\mathrm c0}(\rho_{\mathrm c}/\rho_{\mathrm c0})
\end{equation}
respectively, where $\rho_{\mathrm c0}$, $p_{\mathrm c0}$, and $\phi_{\mathrm 0}$ are the coronal density, gas pressure and potential at $(r,z)=(r_0,0)$, respectively, and $c_{\mathrm{sc}}$ is the constant sound speed in the coronae.
The pressure ratio of corona to disk is $p_{\mathrm c0}/p_{\mathrm 0}\sim0.18$ at $(r,z)=(r_0,0)$
A weak, purely poloidal magnetic field is embedded in the initial disk.
The initial magnetic field distribution is described in terms of the $\varphi$-component of the vector potential $A_\varphi$, which is assumed to be proportional to the density ($A_{\varphi}\propto\rho$) and $r$.
The strength of the magnetic field is parametrized by the ratio of the gas pressure to the magnetic pressure (plasma $\beta$) that is assumed to be constant in the initial disk. 
Table~\ref{tab:param} shows the parameters adopted in the simulation we report in this paper.
Figure~\ref{fig:init} shows the distribution of initial gas density (upper left), gas pressure (upper right), temperature (lower left), and plasma-$\beta$ (lower right) in the poloidal plane at $t=0$.
%
%We normalize the lengths, velocities, and densities  by the Schwarzschild radius, speed of light, and maximum density of the initial disk, $\rho_\mathrm{0}$,  respectively. The unit time is $t_0=r_{\rm s}/c=100\ M_{\rm BH}/(10^7M_\odot)\ {\rm s}$. 
%
The half thickness of the disk is $\sim5r_{\mathrm s}$, and the inner edge of the disk is $r\sim 25r_{\mathrm s}.$
The minimum temperature of the disk is $T\sim10^8\ \mathrm K$.
%
%Outside the disk, we assumed a spherical, isothermal halo with temperature $T\sim10^{12}~\mathrm{K}$.
%
These initial conditions are the same as \citet{kato+2004} \citep[see also][]{hawley+balbus2002}. 
%

%%resolution
The computational domain of our simulation is $0 \le r < 274r_{\rm s}$, $0 \le \varphi < 2\pi$, and $|z| < 233r_{\rm s}$; and the number of grid points is $(n_{\rm r},n_\varphi,n_{\rm z}) = (592,32,704)$. Grid spacing is $0.1r_{\rm s}$  in the radial and vertical directions when $r < 20r_{\rm s}$ and $|z| < 5r_{\rm s}$ and increases outside the region. 
The absorbing boundary condition is imposed at $R=2r_{\rm s}$, and the outer boundaries are free boundaries where waves can be transmitted.
%

%%%%
\section{Results} \label{sec:results}
We avoided including radiative cooling until the disk is heated by accretion.
We carried out numerical simulations without including radiative cooling until RIAF is formed.
In Section 3.1, we present the results of such simulations.
After quasi-steady RIAF is formed, we specify the initial density of the disk $\rho_0$ and carry out simulations considering the radiation process.
When the density exceeds the upper limit for the existence of RIAF, cooling instability grows.
The results of such simulations are presented in Section 3.2.
We note that in AGN, the density of RIAF increases as the mass accretion rate from the outer cold disk increases.
Cooling instability grows in local regions where the density exceeds the threshold.
In Section 3.3, we show that QPOs are excited as a result of the growth of the cooling instability.
%In our simulations presented below, when we specify $\rho_0$ exceeding the upper limit for RIAF, cooling instability will grow and dense blobs formed in RIAF.}
%

%
\subsection{Formation of MHD accretion flow}
During the time range $0<t<t_{\mathrm c}=1.58\times 10^{4}t_0$, which equals seven Keplerian rotation periods at the density maximum of the initial disk at $r = r_0$, we ignored all radiation processes and followed the evolution of the nonradiative accretion flow until quasi-steady state is attained as a result of the growth of the MRI.
\begin{figure*}
\plotone{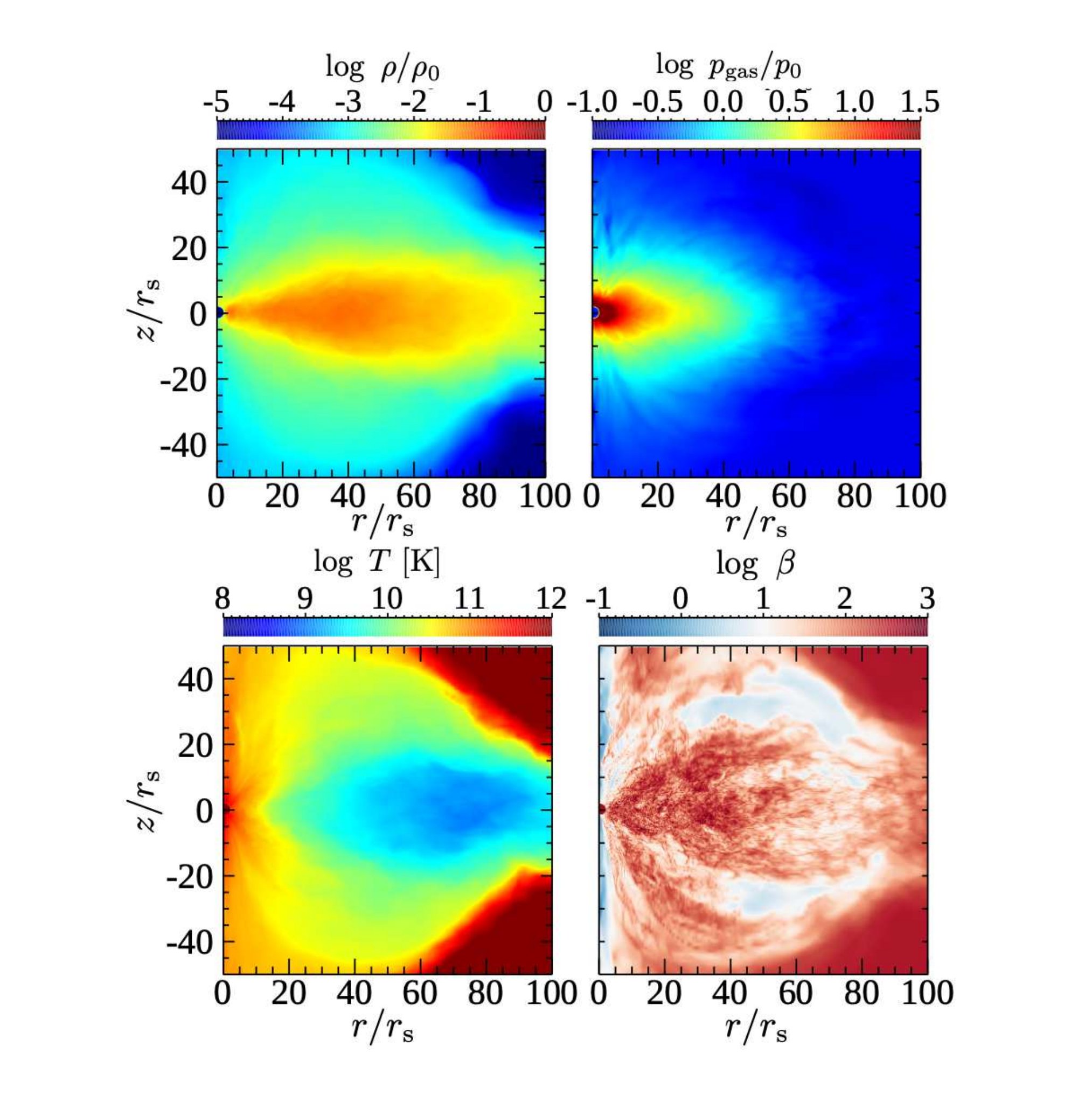}
\caption{Azimuthally averaged distributions of density (upper left), gas pressure (upper right),  gas temperature (lower left), and plasma $\beta$ (lower right) in the poloidal plane at $t=t_{\mathrm c}$.}
\label{fig:mhd}
\end{figure*}
Figure~\ref{fig:mhd} is the same as Figure~\ref{fig:init} but at $t=t_{\mathrm c}$.
%The right panels of the Figure~\ref{fig:init} show the density and temperature distribution in the poloidal plane at $t=t_{\mathrm c}$.
%
Because of the angular momentum transport by Maxwell stress is enhanced by MRI, the initial disk spreads and the density becomes nearly flat in the radial direction up to $r=80r_{\mathrm s}$.
The disk is heated to $T\gtrsim10^{9}\ \mathrm K$ by releasing gravitational energy and the disk expands in the vertical direction.
Gas pressure is dominant inside the disk, but plasma~$\beta$ becomes low in the disk corona where magnetic fields buoyantly escape from the disk.
\begin{figure*}
\epsscale{1.2}
\plotone{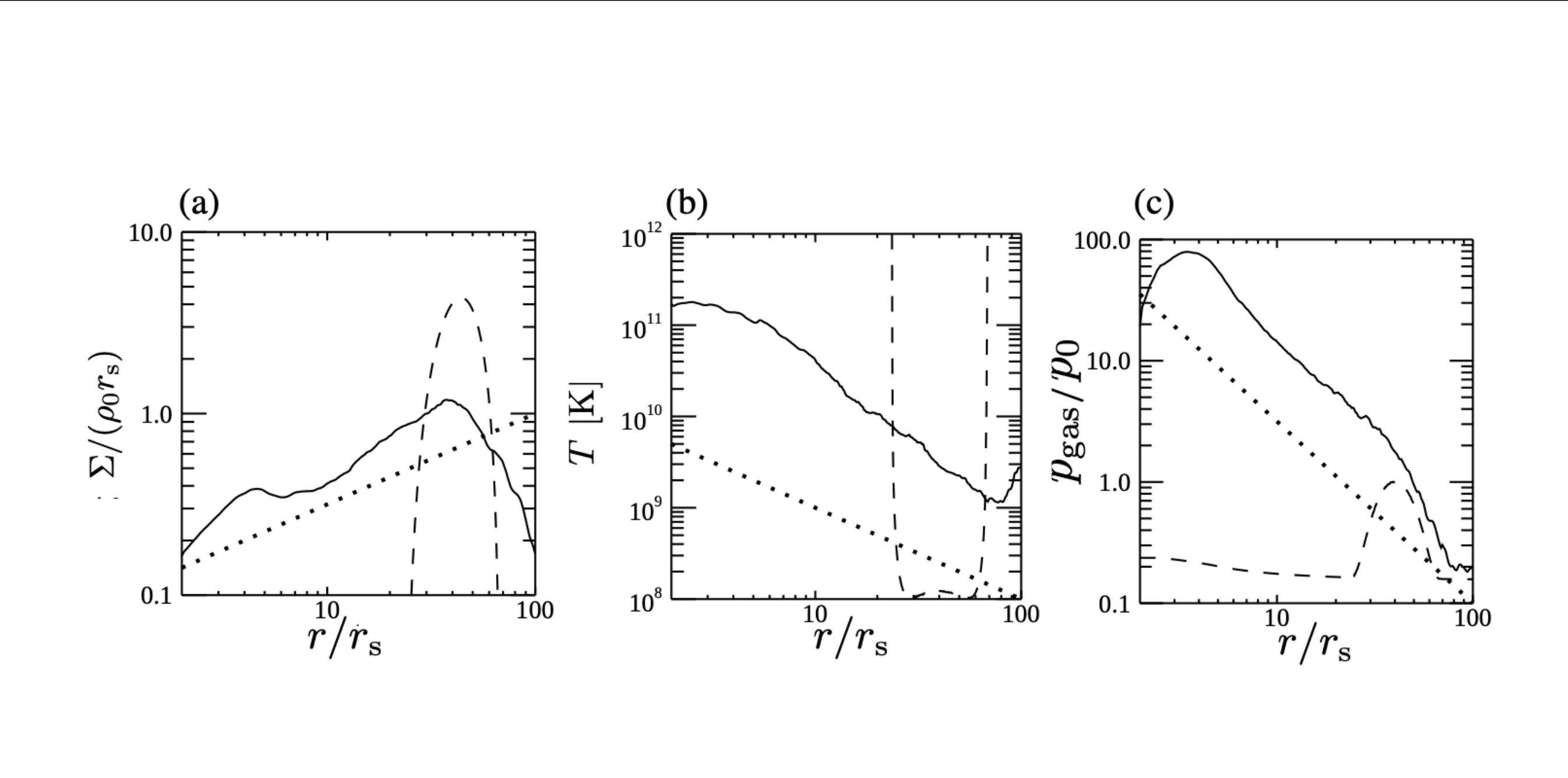}
\caption{Azimuthally averaged radial distribution of (a) surface density, (b) gas temperature and (c) gas pressure at mid-plane ($z=0$). Solid curves show the distribution at $t=t_{\mathrm c}$. Dashed lines show the distribution at $t=0$. Dotted lines show the radial dependence proportional to $r^{0.5}$, $r^{-1}$, and $r^{-1.5}$.}
\label{fig:radialinit}
\end{figure*}
Figure~\ref{fig:radialinit} shows the radial distribution of surface density, gas temperature, and gas pressure at $t=0$ (dashed curves) and at $t=t_{\mathrm c}$ (solid curves).
The surface density $\Sigma$ is calculated from
\begin{equation}
\Sigma=\int^{10r_{\mathrm s}}_{-10r_{\mathrm s}}\rho dz.
\end{equation}
The radial distribution of the surface density shown in Figure~\ref{fig:radialinit}(a) is nearly proportional to $r^{0.5}$.
This radial dependence ($\Sigma\propto r^{0.5}$) is the same as the convection-dominated accretion flow \citep[CDAF, see e.g.,][]{quataert+gruzinov2000}.
\citet{machida+2001} showed that the radial dependence of density and other quantities of nonradiative MHD accretion flow are almost the same as CDAF in which $\rho\propto r^{-0.5},\ T\propto r^{-1}$ and $p_{\mathrm{gas}}\propto r^{-1.5}$. 
The radial distribution of the temperature (Figure~\ref{fig:radialinit}(b)) is slightly steeper than the CDAF solution. % because the angular momentum distribution is nearly Keplerian.
%
%The radial distribution of the gas pressure (Figure~\ref{fig:radialinit} (c)) is proportional to $r^{-1.5}$ and consistent with CDAF solution.
%
Inside the radius of the density maximum of the initial disk ($r<r_0$), the accretion flow is approaching the solution of nonradiative MHD accretion flows.
%
%In this paper, we define the state which the radial dependence of the various quantities close to the CDAF solution even the mass accretion rate is not constant is quasi-steady state.}
%Note that the net mass accretion rate is constant in the steady state solution but the mass accretion rate at $t=t_{\mathrm c}$ constant up to $r\sim20r_{\mathrm s}$ but the radial dependence of quantities is close to steady state solution.}
%
We should note, however, that a small enhancement of surface density remains around $r=r_0$.
%
%This density enhancement is smaller to the initial condition of the 1-dimensional simulation of the evolution of the density enhancement of RIAF reported by \citet{manmoto+1997}, in which they assumed that density enhancement around $r=50r_{\mathrm s}$ with radial wavelength $\lambda\sim10r_{\mathrm s}$.
%
%They followed the propagation of the density enhancement without taking into account the radiative cooling.
%
  
%
%In this paper, we study the evolution of the density enhancement when the surface density exceeds the upper limit for the existence of RIAF by considering the radiative cooling.
%
In this paper, we study the evolution of the overdense region in RIAF when the surface density exceeds the upper limit for RIAF.
We assume that the origin of the density enhancement is the increase of the mass accretion rate from the outer region (see Figure~\ref{fig:schematic}(a)).
%

%%%
%The surface density have small hump around the initial torus center.
%%
%This is slightly higher than RIAF solution ($\Sigma\propto r^0$).
%%
%The distribution of the surface density at $t=1.58\times 10^4t_0$ is similar to the \citet{manmoto+1997} in which they assumed that density enhancement around $r=50r_{\mathrm s}$ with radial wavelength $\lambda\sim10r_{\mathrm s}$ infalls with accretion.
%%
%The temperature is heated up to $T\sim10^{9-10}\ \mathrm K$ by the dissipation of the magnetic field.
%%
%The gas temperature is also different from RIAF solution ($T\propto r^{-1}$).
%%
%This is due to the initial torus rotates Keplerian and the rotation is nearly Kepler at $t=1.58\times10^4t_0$.
%%
%The radial velocity is close to RIAF solution.
%%
%%%

%
We simulate the evolution of the MHD accretion flow when the surface density of the density enhancement exceeds the upper limit for the existence of RIAF.
We choose the maximum density $\rho_\mathrm{0}=2\times10^{-11}~\mathrm{g}/\mathrm{cm}^3$ such that the mass accretion rate of the quasi-steady nonradiative accretion flow is $0.1\dot{M}_\mathrm{Edd}$ at $r=10r_{\rm s}$.
Because this accretion rate exceeds the upper limit for RIAF, radiative cooling exceeds heating.
\begin{figure*}
%\plotone{figure/snapshot.eps}
\epsscale{.8}
\plotone{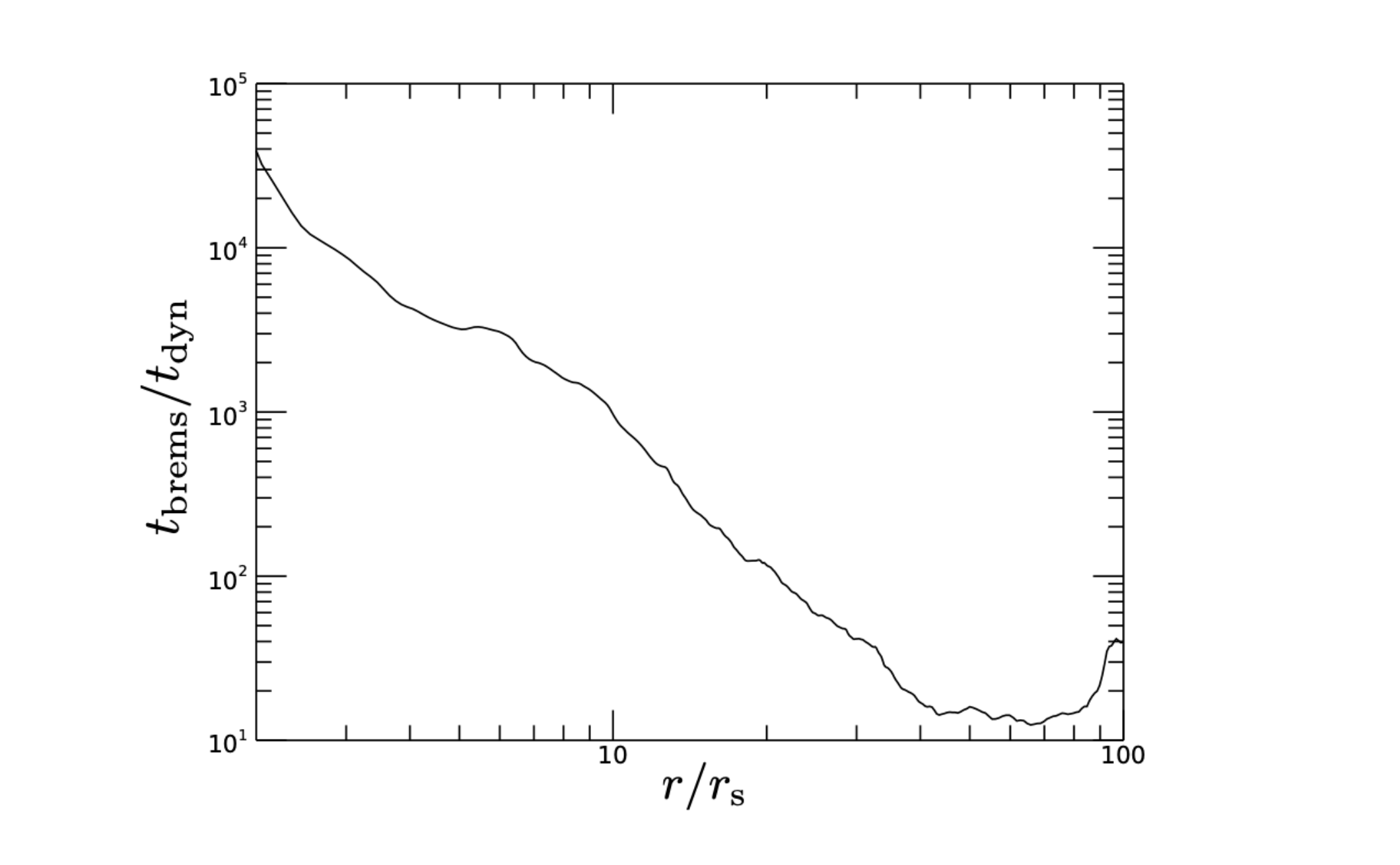}
\caption{Azimuthally and vertically ($|z|<0.5r_{\mathrm s}$) averaged radial distribution of the ratio of the bremsstrahlung cooling time ($t_{\mathrm{brems}}$) and dynamical time ($t_{\mathrm{dyn}}$) at $t=t_{\mathrm c}$.
\label{fig:inittime}}
\end{figure*}
Figure~\ref{fig:inittime} shows the azimuthally and vertically ($|z|<0.5r_{\mathrm s}$) averaged radial distribution of bremsstrahlung cooling time $t_{\mathrm{brems}}$ and dynamical time $t_{\mathrm{dyn}}$ computed using
\begin{equation}
	t_{\mathrm{brems}} = \frac{p_{\mathrm{gas}}/(\gamma-1)}{\rho\kappa_{\mathrm{ff}}a_{\mathrm r}cT^4},
\end{equation}
\begin{equation}
	t_{\mathrm{dyn}} = r/v_{\varphi},
\end{equation}
respectively, where $a_{\mathrm r}$ is the radiation constant.
\textbf{
The cooling time scale is much longer than the dynamical time scale in the inner region ($r<20r_{\mathrm s}$).
%
%Meanwhile, it is $\sim10$ times larger than the dynamical timescale in the region $30r_{\mathrm s}<r<80r_{\mathrm s}$ and our simulation include more than $\sim10$ times dynamical timescale.
%
Meanwhile, the cooling time scale is about $10t_{\mathrm{dyn}}$ in the outer region ($30r_{\mathrm s}<r<80r_{\mathrm s}$).
Because we cover more than $100t_{\mathrm{dyn}}$ at $r=40r_{\mathrm s}$ in our simulation, we expect that cooling instability will grow in this region.
}
%
%Around the initial density maximum ($30r_{\mathrm s}<r<60r_{\mathrm s}$), the cooling time scale of bremsstrahlung emission is around $2t_{\mathrm{dyn}}$ so that we carry out simulation longer than this time scale. 
%
%On the other hand, cooling time scale is much longer than the rotation time scale in the inner region ($r<25r_{\mathrm s}$).
%

%Subsequently, we set the maximum density of the initial torus as $\rho_\mathrm{0}=2\times10^{-11}~\mathrm{g}/\mathrm{cm}^3$ 
%and carried out simulation with radiation processes.
%
%n the beginning  of the RMHD simulation, the density close to the mid-plane is $\rho(z=0) \approx 10^{-5}\ \mathrm{g}/\mathrm{cm}^{3}$ and accretion rate in the vicinity of the black hole is $\dot{M}\approx 0.1~\dot{M}_\mathrm{Edd}$, where $\dot{M}_\mathrm{Edd}=L_\mathrm{Edd}/c^2$ is the Eddington accretion rate. 

\subsection{Results of the radiation MHD simulation}
In this section, we present the results of the RMHD simulation performed by applying the RMHD code CANS+R.
In the simulation, radiation energy density $E_{\rm r}$ at $t=t_{\mathrm c}$ is set to a negligibly small value ($E_{\rm r} = 10^{-20}~\mathrm{erg/cm^3}$) in the whole region.
%

%%%
%\subsection{Coexisting a hot inner region and a cool outer region in the disk}
%%
%{ We present the results of a simulation when $M_{\rm BH}=10^7M_\odot$.}
%{ We choose the maximum density
%%Numerical simulations taking into account the radiative cooling are carried out by specifying the density 
%$\rho_\mathrm{0}=2\times10^{-11}~\mathrm{g}/\mathrm{cm}^3$ such that the mass accretion
%rate of the  quasi-steady nonradiative accretion flow is $0.1\,\dot{M}_\mathrm{Edd}$ at $r=10r_{\rm s}$.
%%
%Since this accretion rate exceeds the upper limit for RIAF, radiative cooling becomes dominant.}

%When the mass accretion rate becomes larger than $0.1\,\dot{M}_\mathrm{Edd}$,  a cool region starts to appear due to radiative cooling.
%%
\begin{figure*}
%\plotone{figure/snapshot.eps}
\epsscale{1.3}
\plotone{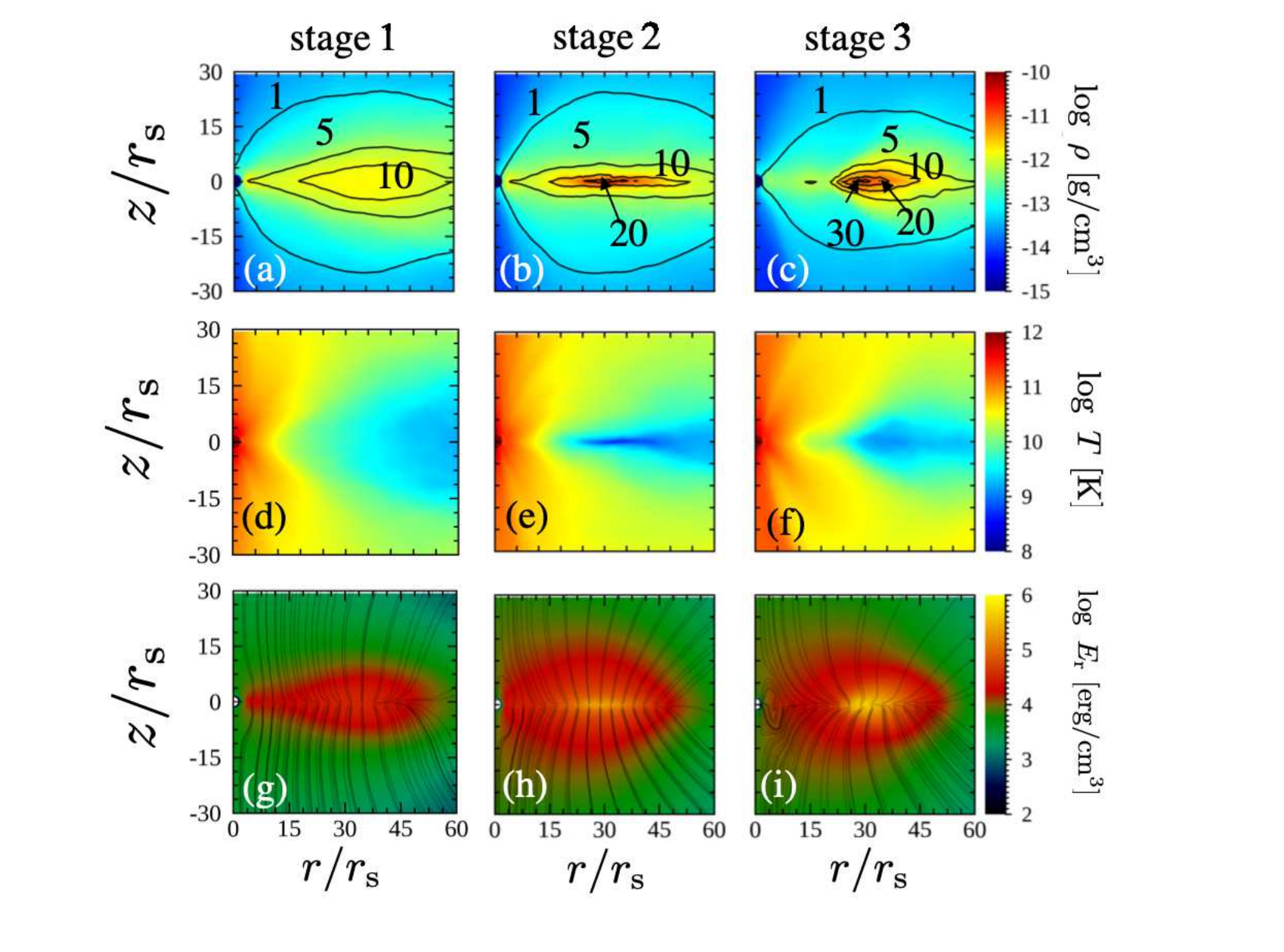}
\caption{Azimuthally averaged gas density, gas temperature, and radiation energy density in the radiation frame. The quantities are averaged over the time range $1.58\times10^4t_0<t<1.78\times10^4t_0$ (stage~1), $3.05\times10^4t_0<t<3.25\times10^4t_0$ (stage~2) and $3.90\times10^4t_0<t<4.10\times10^4t_0$ (stage~3). Contours in (a), (b) and (c) show $\tau_{\mathrm{es}}=1, 5, 10, 20, 30$ and solid curves in (g), (h) and (i) show the streamlines of the radiation energy flux in the poloidal plane.
\label{fig:snapshots}}
\end{figure*}
Figure~\ref{fig:snapshots} shows the distribution of gas density, gas temperature, and radiation energy density in the poloidal plane averaged over azimuthal direction and time.
{We pick up numerical results at three epochs.
Stage 1 ($1.58\times10^4t_0<t<1.78\times10^4t_0$) is the stage just after the radiative process is included.
Stage 2 ($3.05\times10^4t_0<t<3.25\times10^4t_0$) is the stage when the cool region is formed.
%
%Stage 3 ($3.90\times10^4t_0<t<4.10\times10^4t_0$) corresponds to the later stage.
Stage 3 ($3.90\times10^4t_0<t<4.10\times10^4t_0$) is the stage when radiation pressure becomes comparable to gas pressure.}
A remarkable feature of our RMHD simulation is a hybrid radial temperature distribution near the mid-plane (Figure~\ref{fig:snapshots}(d), (e), and (f)).
In the inner region ($r<25r_{\mathrm s}$), hot, low-density, RIAF-like accretion flow similar to those in previous nonradiative MHD simulations is sustained.
In contrast, in the outer region ($r > 25r_{\rm s}$), the disk shrinks in the vertical direction, and a dense and cool disk extending to $r \sim 45r_{\rm s}$ appears (stage~2 and stage~3). 
Figure~\ref{fig:snapshots}(e) and (f) shows that a cool region is formed by radiative cooling around the small density enhancement at $t=t_{\mathrm c}$. %\citep{manmoto+1997}.
The innermost region of the cool disk extends down to $r=25r_{\mathrm s}$.
{%In this region, radiation pressure is dominant, and therefore the cool disk is supported by radiation pressure gradient force.
%{In this region, radiation pressure comparable to the sum of magnetic pressure and gas pressure. Radiation pressure is dominant in the blue region in  Figure~\ref{fig:snapshots} (h). Magnetic pressure is comparable to gas pressure in the cool region, and the region contains low-$\beta$ filaments}.
{%The radiation pressure is comparable to the sum of the magnetic and gas pressures. Deep inside the cool disk, 
%The radiation pressure dominates the gas and magnetic pressures in this region (blue region in Figure~\ref{fig:snapshots} (h)). 
%The magnetic pressure is the second highest following to the radiation pressure. The low beta filaments appears inside the cool disk.}
%
Figure~\ref{fig:snapshots} (b) and (c) shows that the equatorial region of the cool disk is optically thick for Thomson scattering ($\tau_{\mathrm{es}} \gtrsim 10$), where 
\begin{equation}
\tau_{\rm es}=-\int_{+50r_{\mathrm s}}^0 \rho\kappa_{\mathrm {es}}dz.
\end{equation}
{Thomson optical depth around the mid-plane in stage~3 becomes larger than that in stage~2 (Figure~\ref{fig:snapshots} (b) and (c))}.
Meanwhile, the disk is optically thin for absorption. The effective optical depth
\begin{equation}
\tau_{\rm eff}=-\int_{+50r_{\mathrm s}}^0 \rho\sqrt{\kappa_{\mathrm{ff}}(\kappa_{\mathrm{ff}}+\kappa_{\mathrm {es}})}dz.
\end{equation}
is $\tau_{\mathrm{eff}}<0.1$ in the whole region.
%
%Solid curves in figure (a) and (d) show contours at $\tau_{\rm es}=$1, 5, 20, and 100.
%
%As shown in Figure \ref{fig:snapshots} (g), most of {the radiation is emitted from the cool, radiation pressure dominant region.}
%
We found that the radiation energy density increases in the cool region. 
Solid curves in Figure~\ref{fig:snapshots}(g), (h), and (i) show the streamlines of radiation flux.
Figure~\ref{fig:snapshots}(h), and (i) clearly shows that most of the radiation is emitted from the cool region.
%
%The fraction of the radiation from the cool region in (i) increases than (h)

%%
\begin{figure*}
\epsscale{1.3}
\plotone{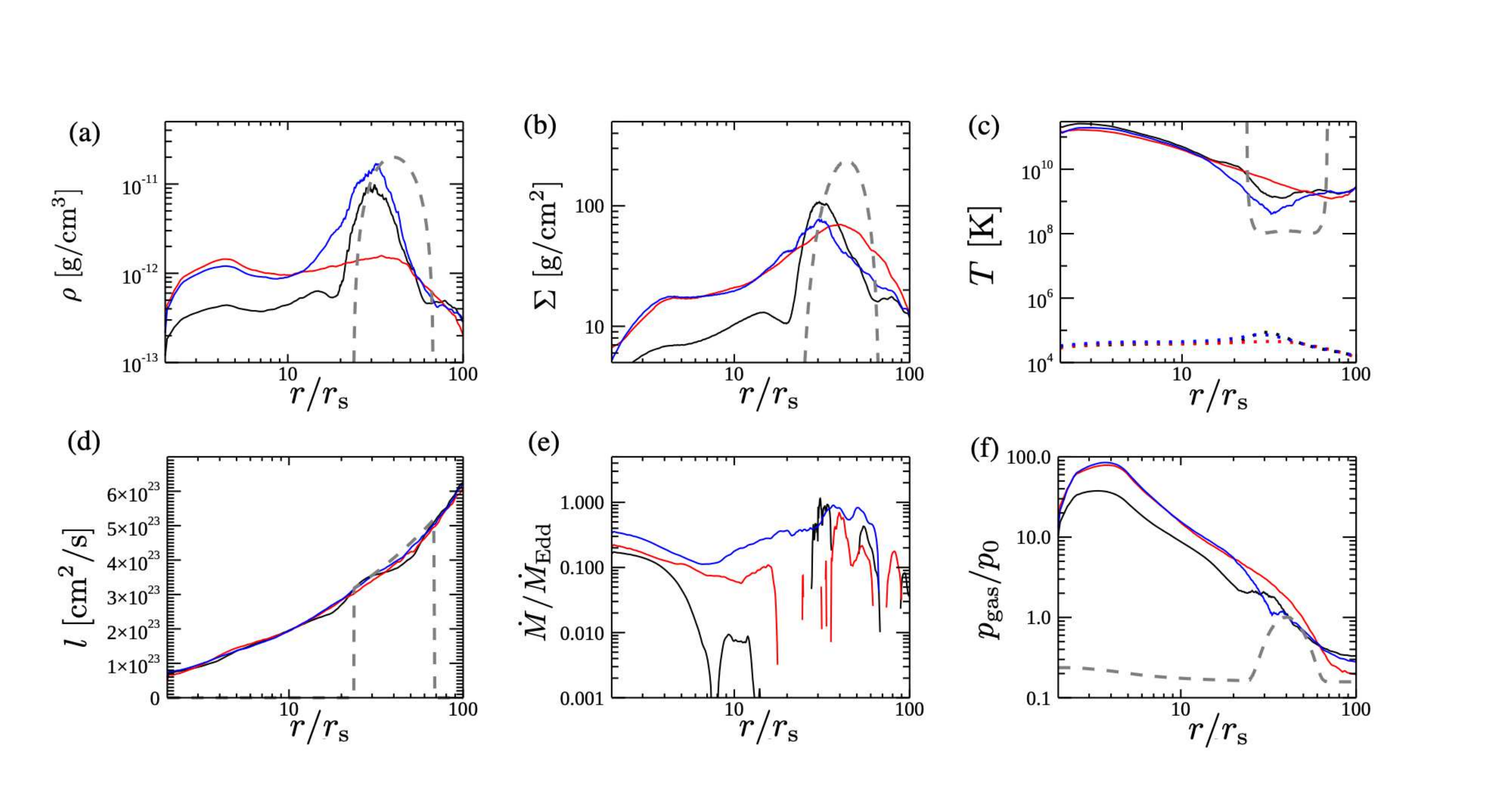}
\caption{Time and azimuthally averaged radial distribution of (a) gas density, (b) surface density (solid), (c) gas temperature (solid) and radiation temperature (dotted), (d) specific angular momentum, (e) the mass accretion rate and (f) the gas pressure at $z=0$. Red, blue and black curves show stage~1, stage~2 and stage~3, respectively. Dashed curves show the distribution at $t=0$. 
\label{fig:radial}}
\end{figure*}
Figure~\ref{fig:radial} shows the azimuthally averaged radial distribution of gas density, surface density, gas temperature and radiation temperature, specific angular momentum, mass accretion rate and radial velocity at stage~1 (red), stage~2 (blue) and stage~3 (black) at $z=0$. 
Figure~\ref{fig:radial}(a) shows that during the interval between stage~1 and stage~2, the density increases by a factor of 10 at $r\sim30r_{\mathrm s}$.  
In contrast, the surface density (the blue curve in Figure~\ref{fig:radial}(b)) does not increase because the disk shrinks in the vertical direction.
The gas temperature around $r\sim30r_{\mathrm s}$ increases between stage~2 and stage~3 because of the dissipation of the magnetic energy.
The radiation temperature is much smaller than the gas temperature even in the cool region.
Figure~\ref{fig:radial}(e) shows the mass accretion rate calculated from
\begin{equation}
\dot{M}=-\int^{2\pi}_0\int^{10r_{\mathrm s}}_{-10r_{\mathrm s}}\rho v_{\mathrm r}dzrd\varphi.
\end{equation}
The mass accretion rate is nearly constant up to $r\sim 20r_{\mathrm s}$ in stage~1 (the red curve) and angular momentum is close to the Keplerian distribution.
Figure~\ref{fig:radial}(f) shows that the radial distribution of the gas pressure is close to the CDAF solution ($p_{\mathrm{gas}}\propto r^{-1.5}$) up to $r\sim80r_{\mathrm s}$ in stage~1 (red curve) and stage~2 (blue curve).
%
%In stage~2, the radial velocity in the cool region is negative, but the mass accretion rate is positive because of the accretion occurred the surface of the cool region.
%The factor change is due to the gas accrete from the outer region.
%
%On the other hand, the angular momentum decrease outside the cool region and the flow into the cool region increase and this is due to the factor change of the surface density.
%
%(b) and (e) The gas temperature and pressure decrease by a factor in the outer region in later stage via the radiative cooling(stage~2 and stage~3).
%
%Furthermore the gas temperature and pressure in the cool region increase stage~2 (the orange curves in (b) and (e)) to stage~3 (the black curves in (b) and (e)).
%
%This is because the region expands in the vertical direction (see Figure~\ref{fig:snapshots}).
%

%%
\begin{figure*}
%\plottwo{figure/cor_mdsig_4.eps}{figure/cor_mdsig_202530.eps}
%\plottwo{f3a.eps}{f3b.eps}
\plotone{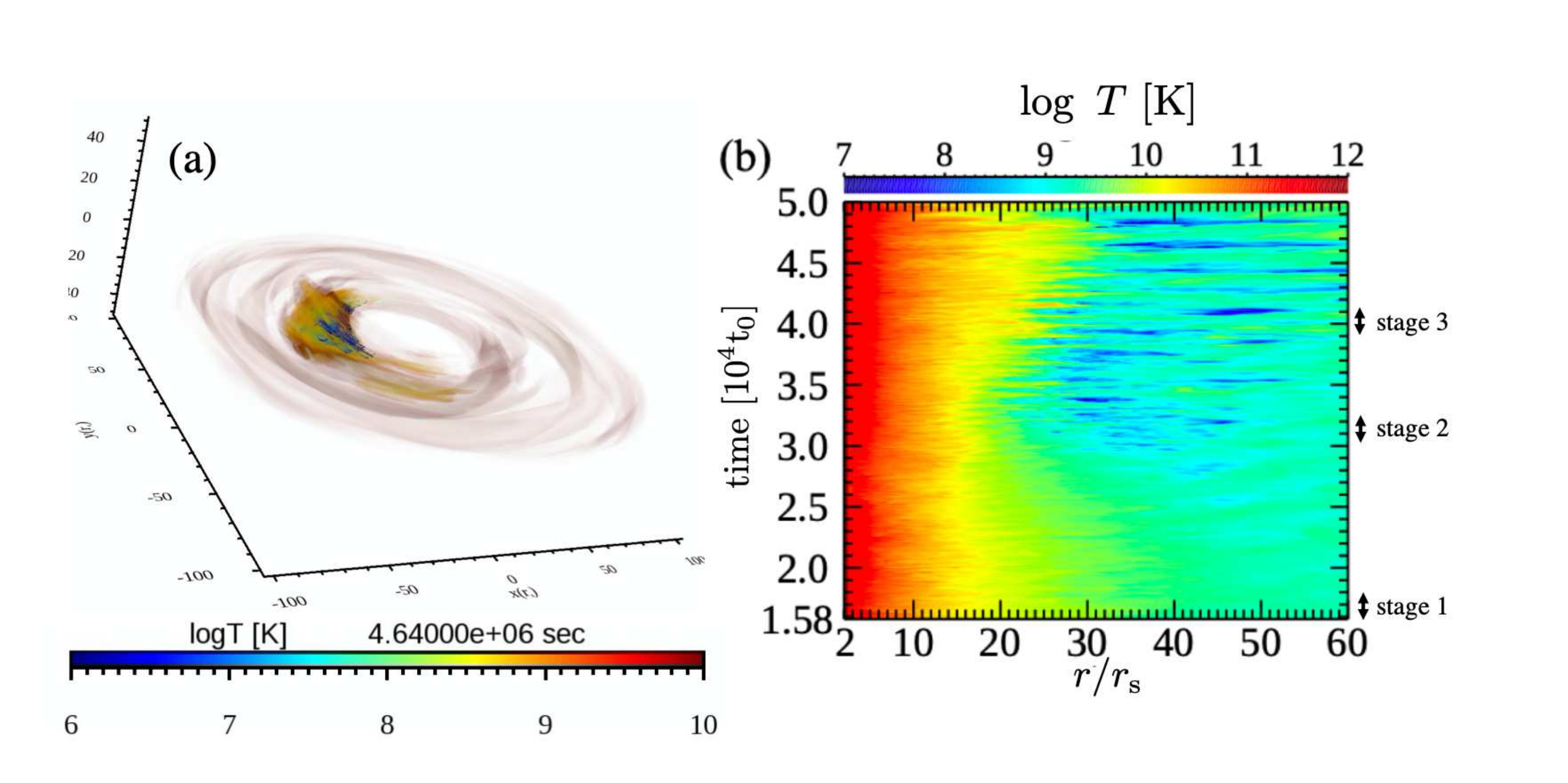}
\caption{(a) Three-dimensional volume rendering image of gas temperature distribution at $t =  4.64\times10^4t_0$. (b) The space--time diagram of gas temperature on the equatorial plane at $\varphi=\pi$.}
\label{fig:temperature}
\end{figure*}
{Figure~\ref{fig:temperature}(a) shows the three-dimensional volume rendering image of the gas temperature distribution at $t=4.64 \times 10^{4}t_0$.
A cool region with a temperature of $T\sim10^7$~K is formed locally.
This temperature is lower than the azimuthally averaged temperature shown in Figure~\ref{fig:snapshots}, and Figure~\ref{fig:radial}(c) because the cool region with temperature $\sim10^7$~K is localized.}
%
%Low temperature ($T \sim 10^{7-8}~{\rm K}$) spiral blob is formed. 
%
Figure~\ref{fig:temperature}(b) shows the space--time plot of the equatorial temperature at $\varphi = \pi$.
As the cooling instability grows, a cool, soft X-ray emitting region appears in the outer disk $(r > 25r_{\mathrm s})$. 
Meanwhile, the region near the black hole ($r < 25r_{\rm s}$) stays hot ($T > 10^{11}~{\rm K})$.
Such a hybrid structure of disk temperature is also shown in general relativistic RMHD simulations \citep{takahashi+2016}.
%
%%Quasi periodic oscillations (QPOs) also appear in the cool outer  region {(see horizontal stripes in Figure~\ref{fig:temperature}} (b)). 
%
%During the stage when $3.0 \times 10^4t_0< t < 4.0 \times 10^4t_0$, the oscillation period is close to the Keplerian rotation period  $\approx 1.1\times10^3t_0$ at $r=25r_{\rm s}$.
%The cool region oscillates quasi-periodically (see horizontal stripes in Figure~\ref{fig:temperature}(b)).
The interface between the hot region and the cool region oscillates quasi-periodically.
The oscillation period is close to the Keplerian rotation period  $\approx 1.0\times10^3t_0$ at $r\sim23r_{\rm s}$.
%
%During the period $3.0 \times 10^4t_0< t < 4.0 \times 10^4t_0$, the oscillation period is close to the Keplerian rotation period  $\sim 1.0\times10^3t_0$ at $r\sim23r_{\rm s}$.
%
%This radius corresponds to the radius of the interface between the hot and cool regions. 
%
%At the stage when $t > 4 \times 10^4t_0$, another oscillation appears in the cool outer region. The oscillation period is $2\times10^3t_0$, which corresponds to the Keplerian period around the outer edge of the cool region at $r\simeq43r_{\mathrm s}$.
In the later stage ($t > 3.5 \times 10^4t_0$), another oscillation appears in the cool outer region (see horizontal stripes in Figure~\ref{fig:temperature}(b)).
The oscillation period is $2.0\times10^3t_0$, which corresponds to the Keplerian period of the cool ($T\sim10^{7-8}$ K) region around $r\simeq40r_{\mathrm s}$.
%
%We would like to present more detailed analysis of the oscillation in section 3.3.
A more detailed analysis of the oscillation is given in Section 3.3.
%

%%%
%\begin{figure*}
%%\plottwo{figure/cor_mdsig_4.eps}{figure/cor_mdsig_202530.eps}
%%\plottwo{f3a.eps}{f3b.eps}
%\plotone{f9.eps}
%\caption{Time evolution of the volume integrated ($20r_{\mathrm s}<r<50r_{\mathrm s}$, $-10r_{\mathrm s}<z<10r_{\mathrm s}$) thermal energy (black), radiation energy (green) and magnetic energy (blue).}
%\label{fig:evene}
%\end{figure*}
%%
%Figure~\ref{fig:evene} shows the time evolution of the volume integrated thermal energy ($E_{\mathrm{gas}}$), radiation energy ($E_{\mathrm{rad}}$) and magnetic energy ($E_{\mathrm{mag}}$) are calculated by
%%
%\begin{equation}
% E_{\mathrm{gas}} = \int^{z_2}_{z_1}\int^{2\pi}_0\int^{r_2}_{r_1} \frac{p_{\mathrm{gas}}}{\gamma -1} dr\,rd\varphi dz,
% \end{equation}
% %
% \begin{equation}
% E_{\mathrm{rad}} = \int^{z_2}_{z_1}\int^{2\pi}_0\int^{r_2}_{r_1} E_{\mathrm r} dr\,rd\varphi dz,
% \end{equation}
%%
%\begin{equation}
% E_{\mathrm{mag}} = \int^{z_2}_{z_1}\int^{2\pi}_0\int^{r_2}_{r_1} B^2 dr\,rd\varphi dz,
% \end{equation}
%respectively, where $r_1=2r_{\mathrm s}$, $r_2=50r_{\mathrm s}$, $z_1=-5r_{\mathrm s}$ and $z_2=5r_{\mathrm s}$, respectively.
% %
%The curve of the thermal energy is multiplied by $0.1$. 
%%
%The thermal energy (black curve) decrease down via the radiative cooling.
%%
%The total magnetic energy decreases after $t=3.0\times10^4t_0$, but increases again after $t=4.5\times10^4t_0$.
%%
%The total radiation energy exceeds the magnetic energy after $t=3.5\times10^4t_0$ and starts to oscillation 
%%
%The oscillation coherent with the time variation of the thermal energy.
%%

%%
\begin{figure*}
\epsscale{1.3}
\plotone{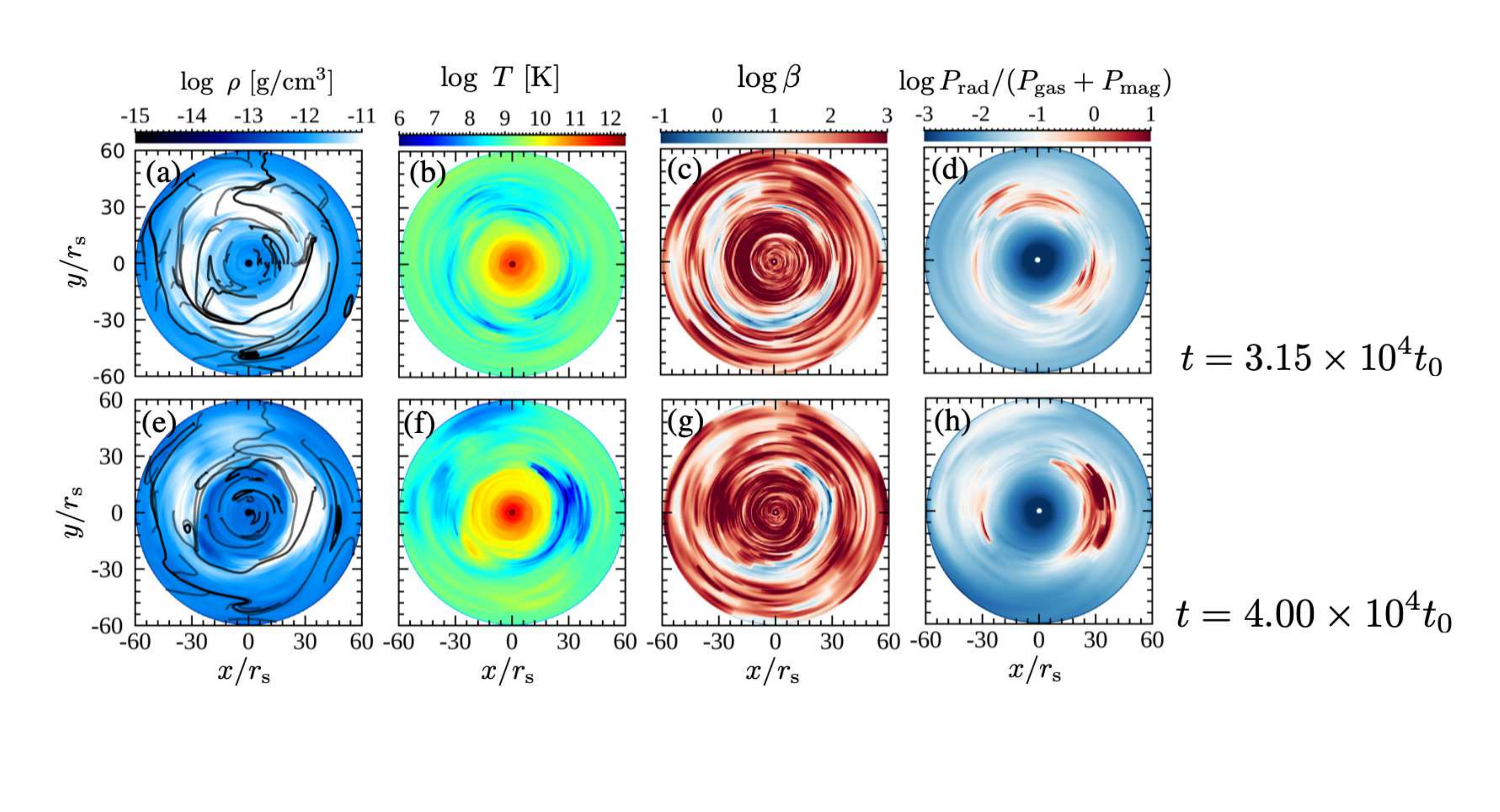}
\caption{Snapshots of the density, gas temperature, plasma $\beta$ and $p_{\mathrm{rad}}/\left(p_{\mathrm{gas}}+p_{\mathrm{mag}}\right)$ in the $xy$ plane averaged in $|z|<0.5r_{\mathrm s}$. The upper panels show the quantities at $t=3.15\times10^4t_0$ and the lower panels show that at $t=4.00\times10^4t_0$. Solid curves in panels (a) and (e) show magnetic field lines.}
\label{fig:snapxy}
\end{figure*}
Figure~\ref{fig:snapxy} shows snapshots of the density, gas temperature, plasma $\beta$ and $p_{\mathrm{rad}}/\left(p_{\mathrm{gas}}+p_{\mathrm{mag}}\right)$ in the $xy$ plane averaged in $|z|<0.5r_{\mathrm s}$.
{Figure~\ref{fig:snapxy}(a), (b) and (c) shows that a dense, low-$\beta$ and cool, soft X-ray emitting region appears outside the inner hot flow ($r>25r_{\mathrm s}$) at $t=3.15\times10^4t_0$. 
%
%In this stage, the cool region dominated by magnetic pressure and radiation pressure.
%
The dense, cool region has a nonaxisymmetric distribution.
%
%{Figure~\ref{fig:snapxy} (b) and (f) show that the tempeature in the cool region is $T\sim10^7~\mathrm{K}$.}
%This is the cause of the azimuthally averaged gas temperature increases in the Figure~\ref{fig:snapshots} (f) and Figure~\ref{fig:radial} (c).
%
Asymmetry becomes more prominent at $t=4.00\times10^4t_0$.
Furthermore, Figure~\ref{fig:snapxy}(h) shows that the radiation pressure becomes dominant in the cool region.}
The effective optical depth of this region is less than unity but radiation is trapped because the Thomson optical depth is large.
%The radiation pressure gradient increase and the cool region expand in the vertical direction compare to the early stage (stage~2 to stage~3).
%
The solid curves in Figure~\ref{fig:snapxy}(a) and (e) show magnetic field lines projected onto the $xy$ plane. 
The magnetic fields are nearly azimuthal in Figure~\ref{fig:snapxy}(a) but deformed into a spiral shape in Figure~\ref{fig:snapxy}(e).
\begin{figure*}
\plotone{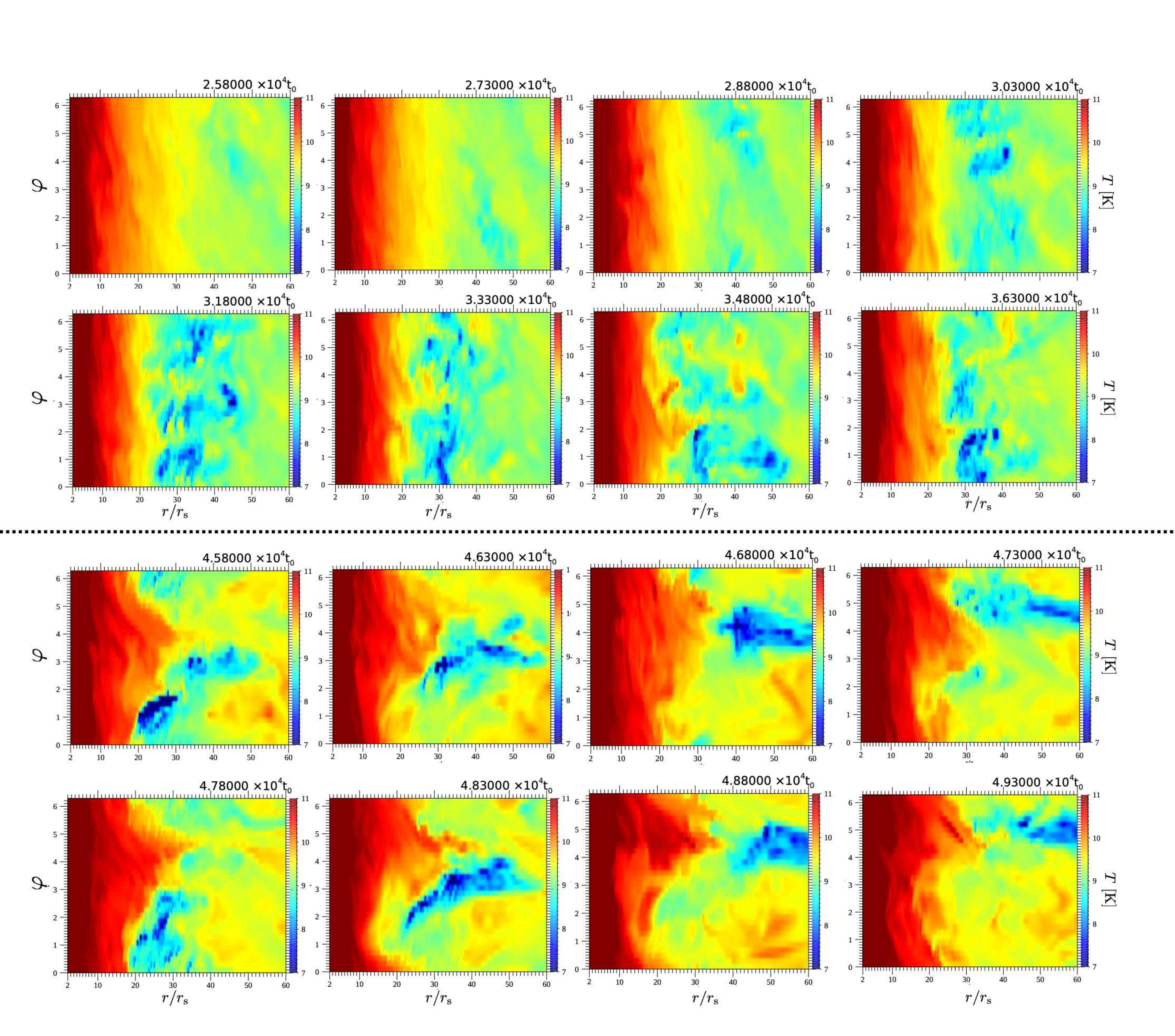}
\caption{The time evolution of the gas temperature distribution in $r-\varphi$ plane averaged in $|z|<0.5r_{\mathrm s}$. The upper eight panels show the onset of the cooling instability and the lower eight panels show the later stage.\label{fig:terphi}}
\end{figure*}
\textbf{
Figure~\ref{fig:terphi} shows the time evolution of the gas temperature distribution in the $r-\varphi$ plane averaged in $|z|<0.5r_{\mathrm s}$.
Panels above the dotted line show the onset of the cooling instability.
Between $t=2.58\times10^4t_0$ and $t=2.88\times10^4t_0$, the temperature gradually decreases from $10^{9.5}$ K to $10^{8.5}$ K around $r\sim45r_{\mathrm s}$.
The temperature distribution has weak nonaxisymmetry.
A cool blob with temperature $T=10^7$ K is formed around $t=3.03\times10^4t_0$.
This cool blob rotates along nearly circular orbit with the local Keplerian rotation speed.
The lower panels in Figure~\ref{fig:terphi} show the temperature distribution in the later stage with a shorter time interval.
The cool region oscillates radially between $r=20r_{\mathrm s}$ and $60r_{\mathrm s}$ with a period around $2.0\times10^3t_0$.
The cool region is not axisymmetric but localized within a finite azimuthal angle.
%
%The inner hot region also shows nonaxisymmetric distribution with $m=1$.
%
In addition, a one-armed pattern grows around the interface between the hot flow and cool disk, and rotates slowly in the azimuthal direction.
%
%More detailed analysis of the oscillation will be given in section 3.3.
}

\begin{figure*}
\epsscale{0.8}
\plotone{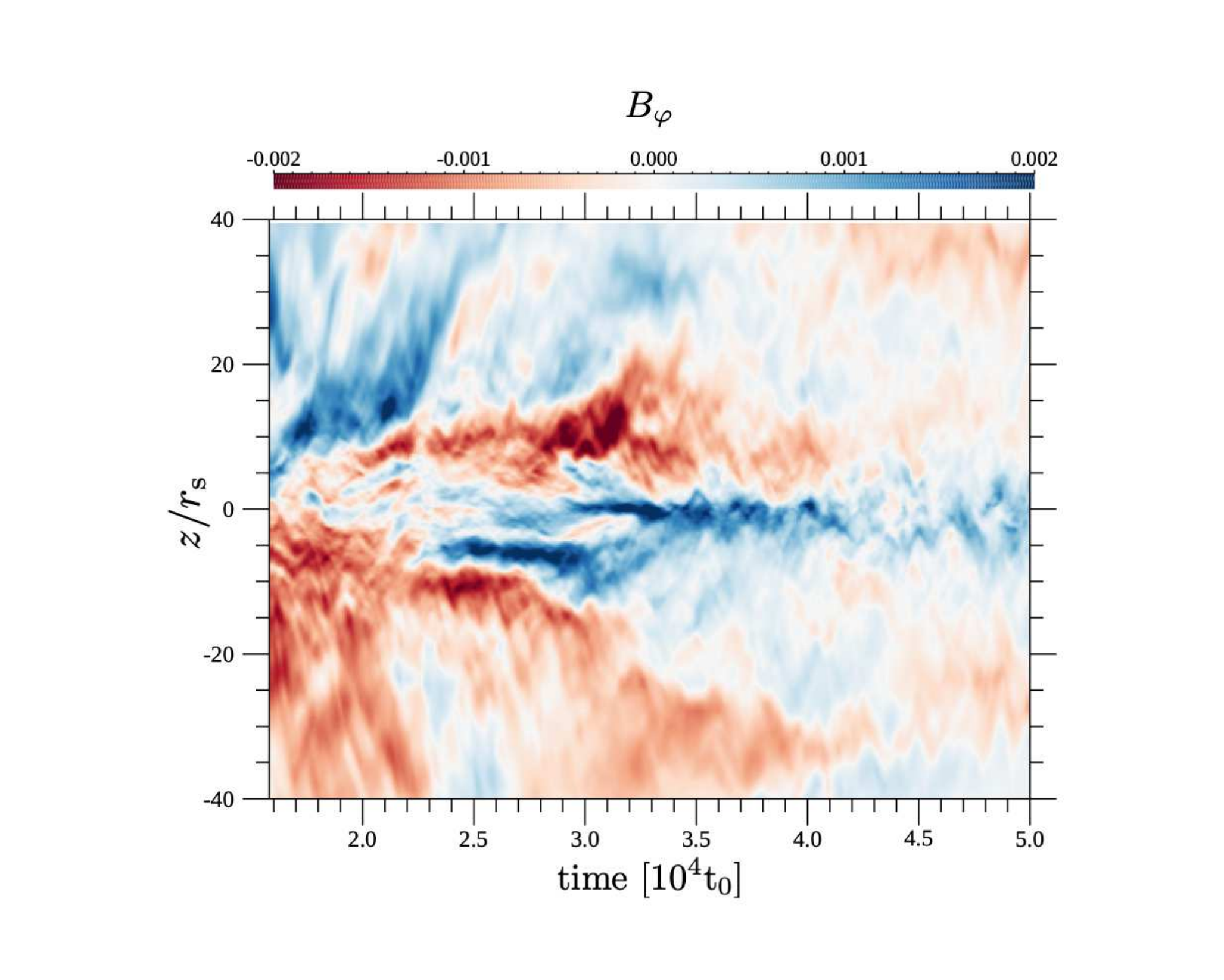}
\caption{Butterfly diagram of azimuthally averaged azimuthal magnetic fields averaged in the radial range $25r_{\mathrm s}<r<35r_{\mathrm s}$.}
\label{fig:butterfly}
\end{figure*}
Figure~\ref{fig:butterfly} shows the butterfly diagram of the azimuthal magnetic field.
The horizontal axis is time, and the vertical axis is height.
The color shows the azimuthal magnetic field averaged in $25r_{\mathrm s}<r<35r_{\mathrm s}$.
The azimuthal magnetic fields buoyantly escape from the disk to the disk corona.
The azimuthal magnetic field at mid-plane increases because of the vertical contraction of the disk.
Because the azimuthal magnetic fields are antisymmetric with respect to the disk mid-plane, they dissipate by magnetic reconnection.
This can be the origin of the heating observed in Figure~\ref{fig:radial}(c).
%
%The magnetic field direction is different in and above the disk and the magnetic energy dissipate and the disk heated up via the magnetic reconection.
%
The azimuthal magnetic field occasionally changes its sign and buoyantly escapes from the disk.
%.

%%
\begin{figure*}
%\plottwo{figure/cor_mdsig_4.eps}{figure/cor_mdsig_202530.eps}
%\plottwo{f3a.eps}{f3b.eps}
\epsscale{0.8}
\plotone{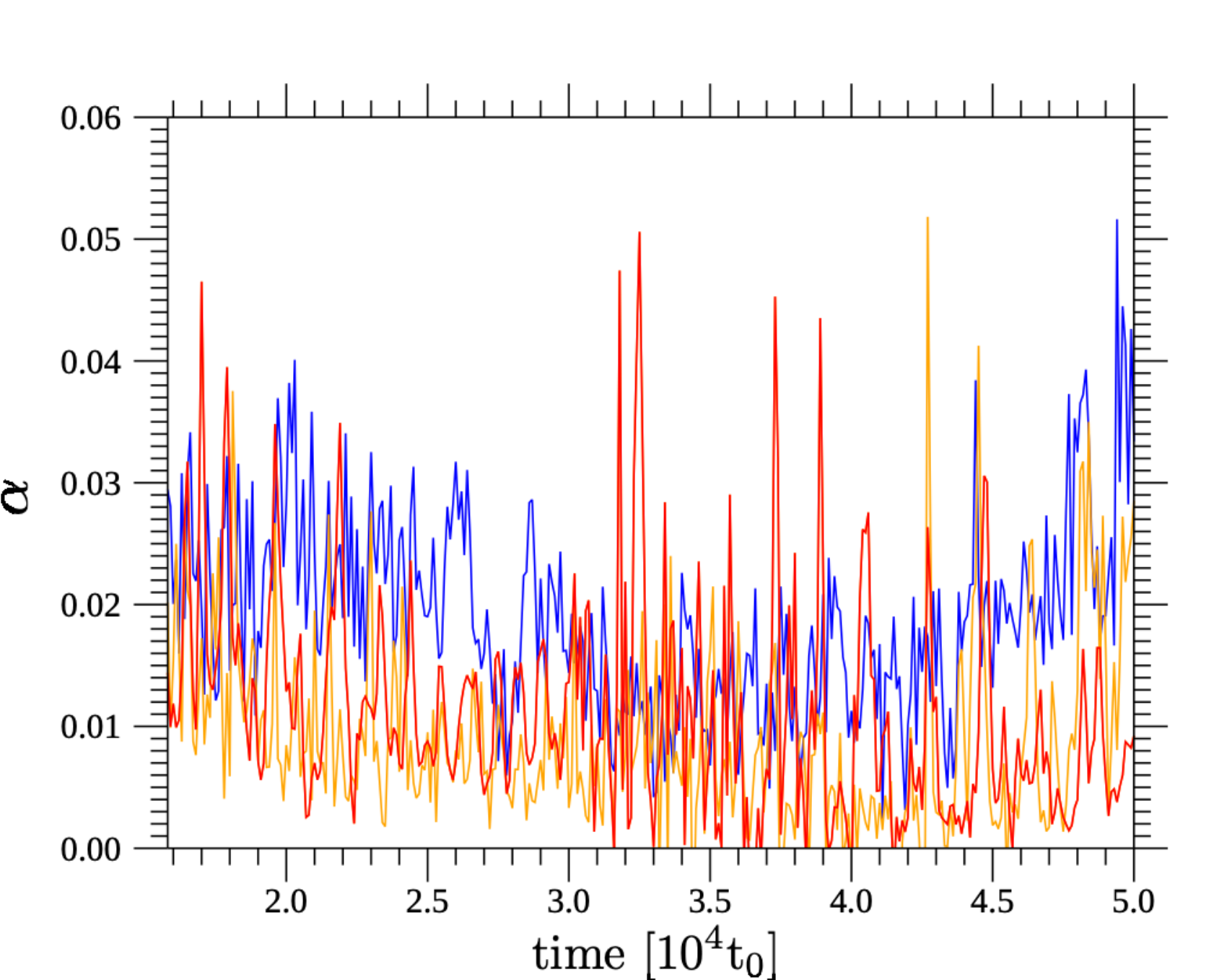}
\caption{Time evolution of $\alpha=-<B_{\mathrm r}B_\varphi>/<p_{\mathrm{gas}}+p_{\mathrm{rad}}>$ averaged over $2r_{\mathrm s}<r<20r_{\mathrm s}$ (blue), $20r_{\mathrm s}<r<25r_{\mathrm s}$ (orange), and $25r_{\mathrm s}<r<50r_{\mathrm s}$ (red), with $|z|<5r_{\mathrm s}$}.
\label{fig:alpha}
\end{figure*}
Figure~\ref{fig:alpha} shows the time evolution of the $\alpha$ parameter \citep{shakura+sunyaev1973}, calculated by,
\begin{equation}
\alpha = -\frac{\left<B_{\mathrm r}B_\varphi\right>}{\left<p_{\mathrm{gas}}+p_{\mathrm{rad}}\right>}
\end{equation}
where $<>$ denotes the volume-weighted average in the cylindrical coordinate and blue, orange, and red show $\alpha$ computed by averaging the region $2r_{\mathrm s}<r<20r_{\mathrm s}$, $20r_{\mathrm s}<r<25r_{\mathrm s}$, and $25r_{\mathrm s}<r<50r_{\mathrm s}$, with $|z|<5r_{\mathrm s}$.
In the inner region (blue curve), $\alpha\sim0.02$ is sustained until $t\sim2.5\times10^4t_0$, when the magnetic flux buoyantly escapes from the disk. %around $t=2.5\times10^4t_0$.
In the outer region (orange and red curve), $\alpha$ increases during the contraction of the disk in $2.7\times10^4t_0<t<3.3\times10^4t_0$ because the azimuthal magnetic field is enhanced in the mid-plane because of vertical contraction.
%
%{After $t>4.0\times10^4t_0$, $\alpha$ decreases due to the magnetic field dissipate inside the disk.}
%Furthermore, $\alpha$ value also oscillates in the outer region, due to the disk repeats the appearance and disappearance of the cool region.
%
In the later stage ($t>4.5\times10^4t_0$), $\alpha$ increases again in the inner region because magnetic fields are regenerated  by the dynamo activity.
The mass accretion is sustained in the whole disk because of the angular momentum transport driven by the Maxwell stress.
%
%\textbf{We should note that the azimuthally averaged radiation pressure is smaller than the azimuthally averaged gas pressure so that the radiation pressure doesn't contribute to the $\alpha$.}

%%%
%\begin{figure*}
%%\plottwo{figure/cor_mdsig_4.eps}{figure/cor_mdsig_202530.eps}
%%\plottwo{f3a.eps}{f3b.eps}
%\plotone{f13.eps}
%\caption{Space--time diagram of the mass outflow rate.}
%\label{fig:mout}
%\end{figure*}
%%%
%Figure~\ref{fig:mout} shows the space--time diagram of the mass out flow rate $\dot{M}_{\mathrm{out}}$ calculated by
%\begin{equation}
%\dot{M}_{\mathrm{out}}(r)=\int\rho v_{\mathrm z}rd\varphi dr.
%\end{equation}
%%
%Around $r\sim10r_{\mathrm s}$, the mass accretion takes place stasionaly.
%%
%This region correspond to the surface of the inner hot flow.
%%
%Around $r\sim20r_{\mathrm s}$, the mass out flow takes place intermittently.
%%
%The mass outflow cause the decrease of the mass accretion rate in the inner region.

%%
\begin{figure*}
%\plottwo{figure/cor_mdsig_4.eps}{figure/cor_mdsig_202530.eps}
%\plottwo{f3a.eps}{f3b.eps}
\plotone{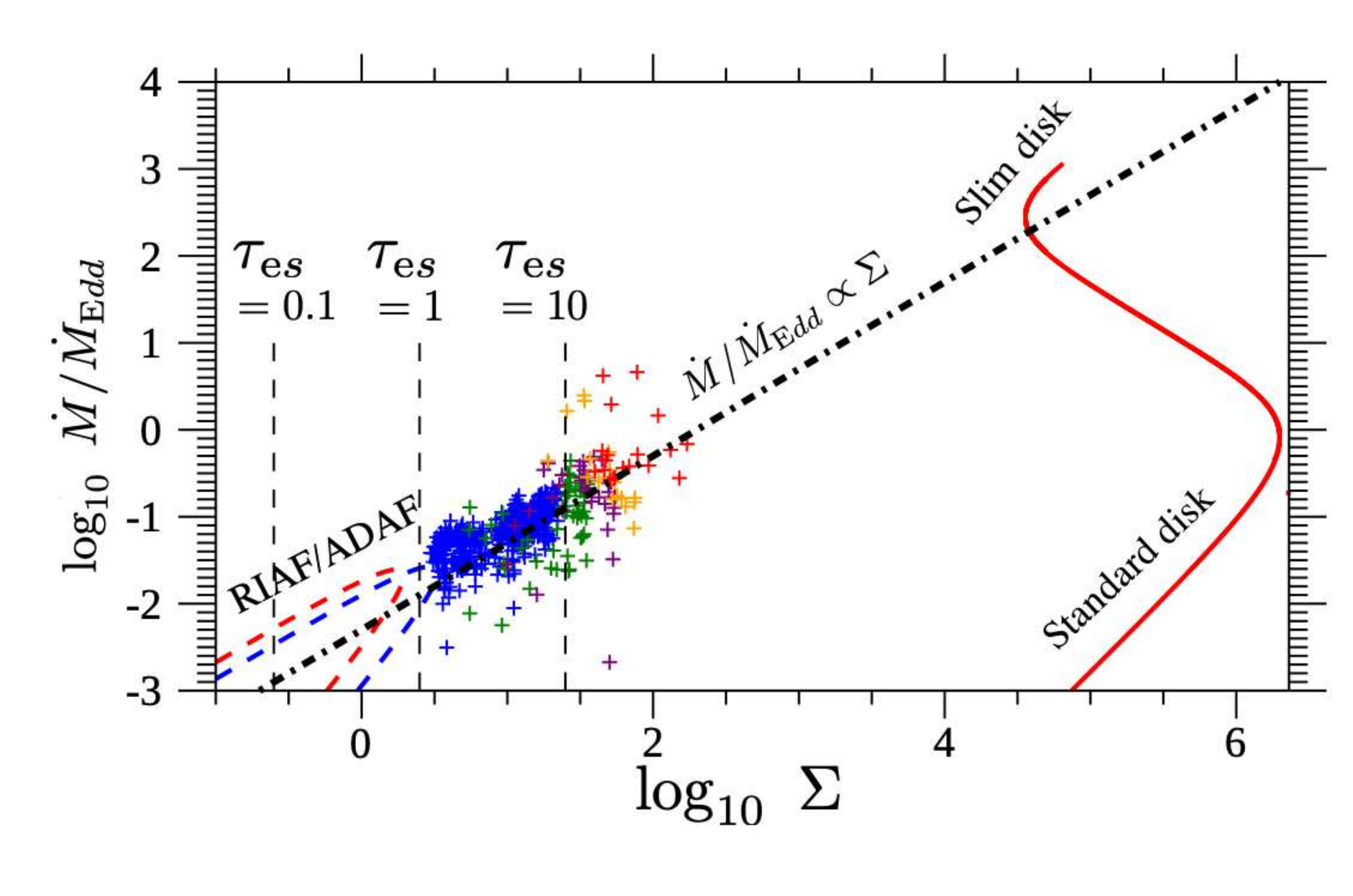}
\caption{Scatter plot of the surface density and  mass accretion rate at the sampling radii $r=5r_{\rm s}$ (blue), $15r_{\rm s}$ (green), $20r_{\rm s}$ (purple), $25r_{\rm s}$ (orange), and $30r_{\rm s}$ (red) and averaged over the Keplerian rotation period at each radius. The dashed blue curve and dashed red curve show equilibrium solutions of optically thin RIAF at $r=5r_{\mathrm s}$ and $r=30r_{\mathrm s}$, and a solid red curve shows an optically thick solution at $r=30r_{\rm s}$ for $\alpha=0.01$ and $M_{\mathrm{BH}}=10^7M_{\odot}$.}
\label{fig:correlation}
\end{figure*}
Figure~\ref{fig:correlation} shows the scatter plot of $\Sigma$ and $\dot{M}$ at various sampling radii.
Dashed curves are thermal equilibrium curves of RIAF at $r=5r_{\mathrm s}$ (blue) and $r=30r_{\mathrm s}$ (red) when the viscosity parameter is $\alpha = 0.01$. 
The dash-dotted line shows the relation $\dot M \propto \Sigma$.
%
%Numerical results in the hot inner region ($r < 25r_\mathrm{s}$: blue, green, and purple) are {consistent with the slope of RIAF solutions.}
Numerical results in the hot inner region ($r<25r_\mathrm{s}$: blue, green, and purple) follow RIAF solutions of $\dot{M}\propto\Sigma$.
%
%The correlation between the surface density $\Sigma$ and the mass accretion rate $\dot{M}$ is close to linear.
%
Note that the accretion rate becomes smaller near the black hole because of the outflows from the hot region.
In contrast, numerical solutions in the cool outer region ($r > 25r_\mathrm{s}$: orange and red) scatter along with the extension of the RIAF solution, and the oscillations occur when the electron scattering optical depth becomes larger than $\sim10$.
%
%, and roughly follows $\dot M \propto \Sigma$ curve.
%
%A gap between the optically thin and thick branch corresponds to the so-called intermediate-type solutions \citep{2001MNRAS.324..119Y, 2012PASJ...64...15O}.
%
%he intermediate-type solutions have been known that there are two distinctive characteristics.
%
Steady  solutions of black hole accretion flow connecting the optically thin RIAF and optically thick standard disk have been obtained by \citet{oda+2009} considering the azimuthal magnetic fields. 
In their solution, radiative cooling is balanced with heating by the dissipation of magnetic energy enhanced by the vertical contraction of the disk because of cooling. 
Furthermore, the disk stays geometrically thicker than the standard model because the magnetic pressure gradient can support the disk \citep[see also][]{jiang+2019sub}. 
We should note, however, that the solutions by \citet{oda+2009} are for stellar-mass black holes.
When the mass of the black hole is $10^7M_{\odot}$, radiation pressure is not negligible when the accretion rate exceeds $1\%$ of the Eddington accretion rate. 
%Figure 3 shows that the disk is partially supported byradiation pressure.
%Our results suggest that the cool region contains both characteristics, because the majority of the cool region is consistent with LHAF and low-$\beta$ filaments also appears in the dense and cool patches in the disk.
%
%Unlike in the hot inner region, the correlation between the surface density $\Sigma$ and the mass accretion rate $\dot{M}$ is slightly difficult to discern in the cool outer  region.
%
%\bf In fact the mass accretion rate becomes a few orders of magnitude larger than the ADAF/RIAF solution when the surface density approaches $10^{2}~\mathrm{g/cm^{2}}$.}
%
%\bf This suggests that some kinds of the optically thin radiative cooling instability take place during the occurrence of changing look (CL) phenomena.}
%

%%
\subsection{Quasi-periodic oscillations excited during the formation of the soft X-ray emitting region}
\begin{figure*}
\epsscale{1.0}
\plotone{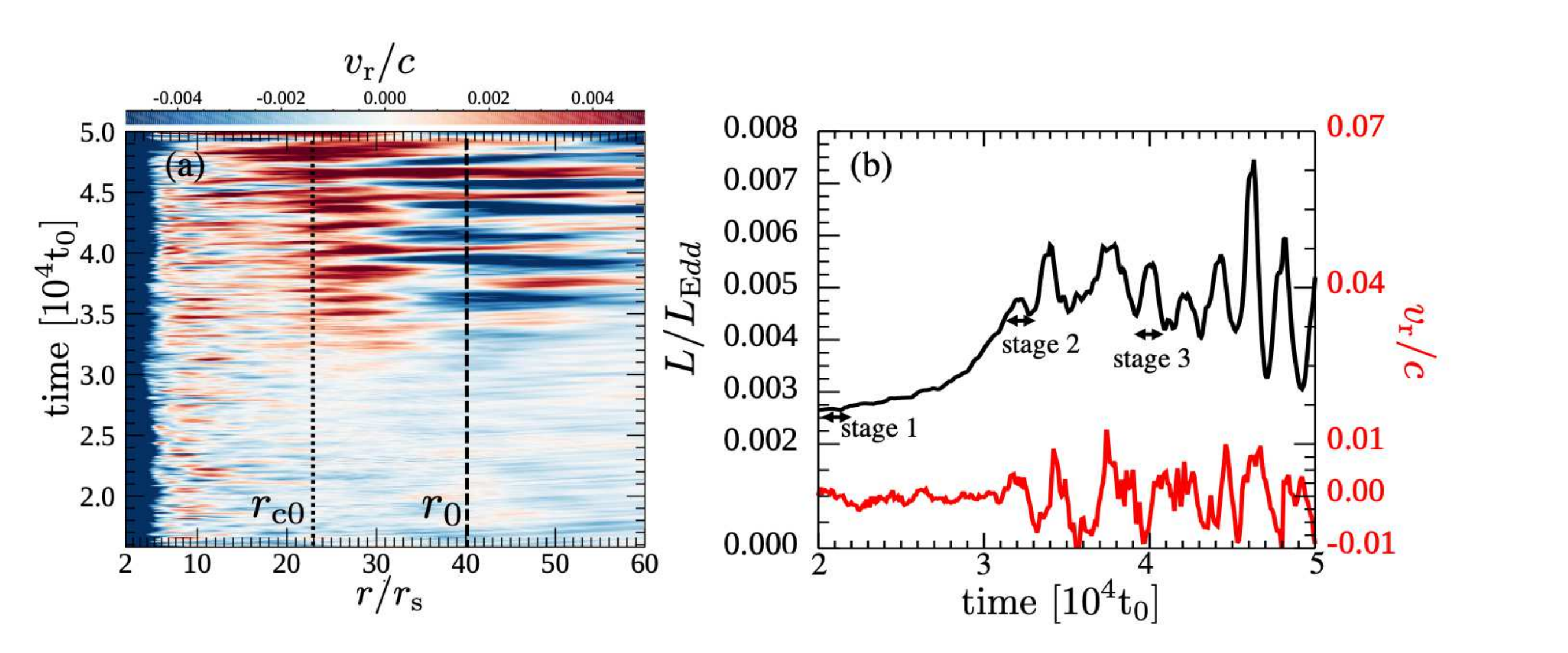}
\caption{(a) Space--time diagram of the radial velocity averaged in the azimuthal direction and $|z|<0.5r_{\mathrm s}$. (b) Disk bolometric luminosity (black) and radial velocity at $z=0$ averaged over $0<\varphi<2\pi$ in $42r_{\mathrm s}<r<44r_{\mathrm s}$ (red). 
The dotted line in (a) indicates the corotation radius $r_{\mathrm c0}$, where the oscillation period equals $1.0\times10^3t_0$ .
The dashed line in (a) indicates the radius of the initial density maximum at $r=r_0$.\label{fig:lum}} %Arrows show the time interval for the time average. \label{fig:lum}}
\end{figure*}
\begin{figure*}
\epsscale{1.2}
\plotone{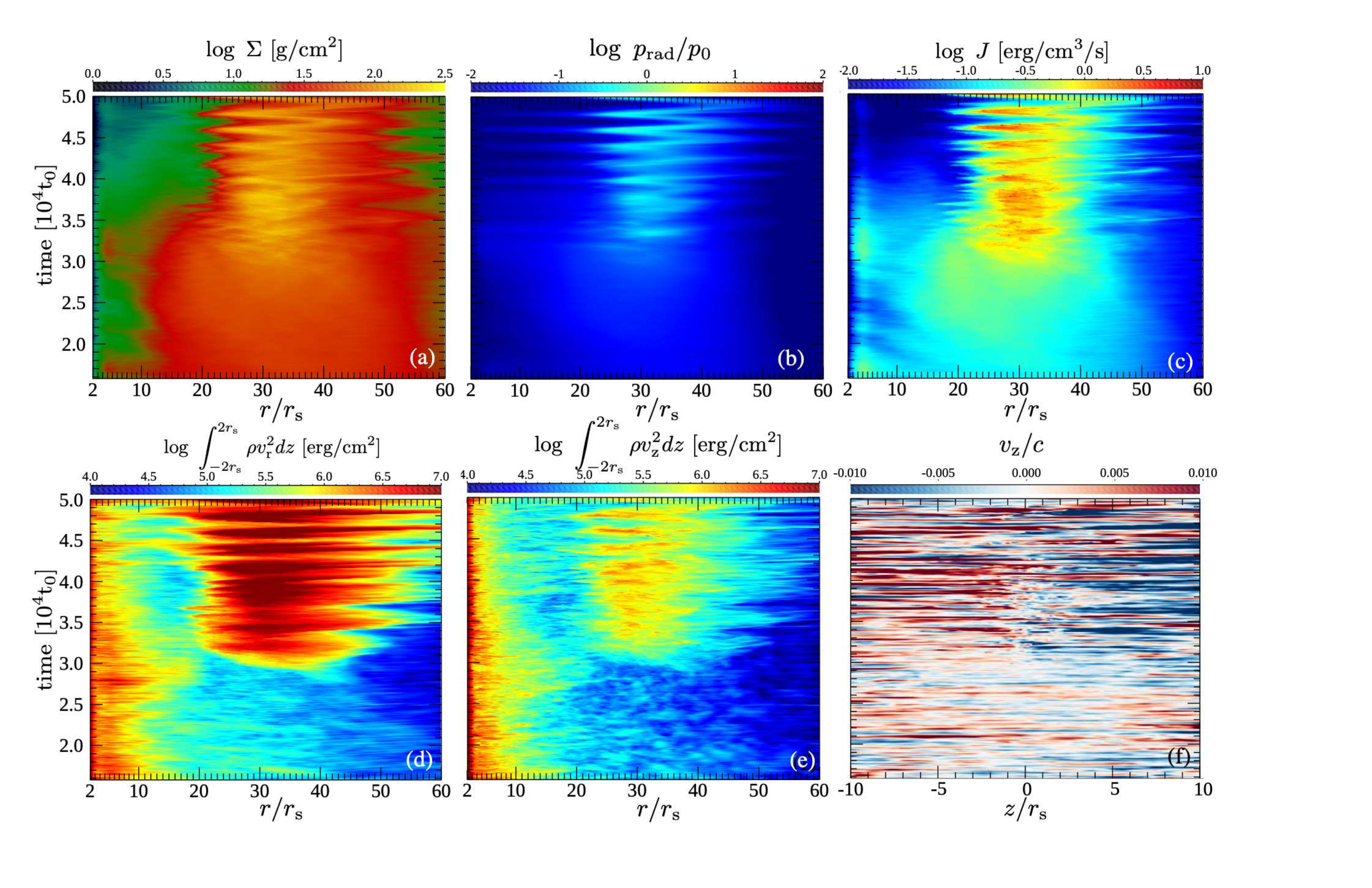}
%\caption{(a) Space--time diagram of the mass accretion rate. Dotted and dashed lines indicate the corotation radius $r_\mathrm{c0}$ and $r_\mathrm{c1}$ where $\bar{\omega} = \omega - m\Omega_\mathrm{K} =  0$. (b) Time variations of the disk bolometric luminosity (black) and the radial velocity at $z=0$ averaged over $0<\varphi<2\pi$ and $42r_{\mathrm s}<r<44r_{\mathrm s}$ (red). Arrows show the time interval for the time average.}
\caption{Space--time diagram of the azimuthally averaged (a) surface density, (b) radiation pressure, and (c) net emissivity $J$ in the $r-$time plane. The radiation pressure and emissivity are averaged in $|z|<0.5r_{\mathrm s}$. (e) and (f) are space--time diagrams of the kinetic energy density in radial and vertical motion integrated in $|z|<2r_{\mathrm s}$. (f) shows a space-time diagram of the azimuthally averaged vertical velocity at $r=30r_{\mathrm s}$ in the $z-$time plane.}
\label{fig:stradi}
\end{figure*}
%%

%\begin{figure*}
%\plotone{v2.eps}
%\caption{Space--time diagram of the azimuthally averaged square of radial and vertical velocity integrated in $|z|<0.5r_{\mathrm s}$. \label{fig:lum}}
%\end{figure*}
%%

%%
\begin{figure*}
\epsscale{1.2}
\plotone{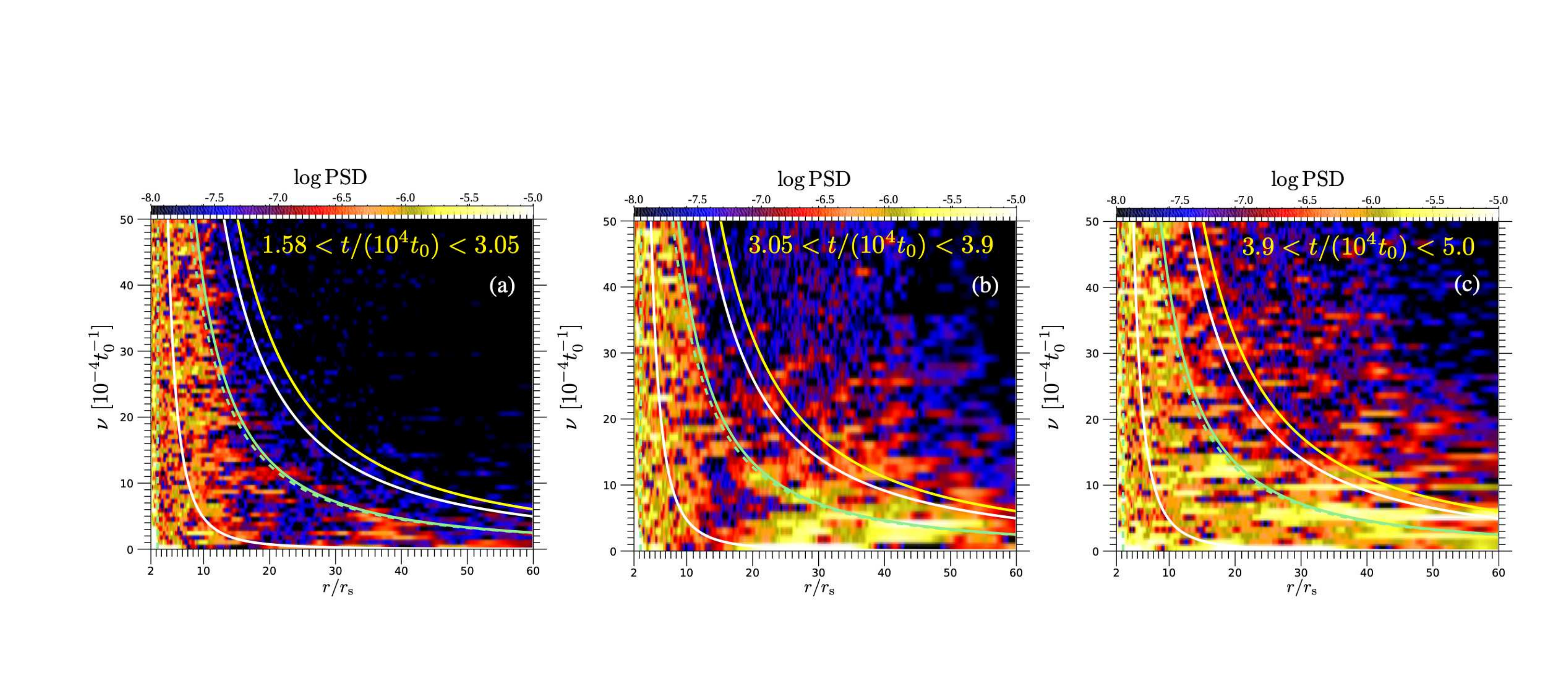}
\caption{The power spectral density (PSD) of the azimuthally averaged radial velocity at mid-plane. The vertical axis is frequency and horizontal axis is radius.  (a), (b), and (c) show the PSD in the time interval in $1.58\times10^4t_0<t<3.05\times10^4t_0, 3.05\times10^4t_0<t<3.9\times10^4t_0$, and $3.9\times10^4t_0<t<5.0\times10^4t_0$, respectively. Green solid and dashed curves are the Keplerian rotation frequency $\Omega_{\mathrm K}/(2\pi)$ and $\kappa/(2\pi)$ where $\kappa$ is the epicyclic frequency. White solid curves show $(\Omega_{\mathrm K}\pm\kappa)/(2\pi)$, and the yellow curve shows $(1+\sqrt{2})\Omega_{\mathrm K}/(2\pi)$.}
\label{fig:psd}
\end{figure*}
\textbf{
One of the most remarkable findings is that the disk oscillates quasi-periodically as the soft X-ray emitting region is formed outside the inner hot RIAF.
Figure~\ref{fig:lum} (a) shows the space--time diagram of the azimuthally averaged radial velocity in $|z|<0.5r_{\mathrm s}$.
The radial oscillation becomes prominent around $r\sim23r_{\mathrm s}$ in $3.05\times10^4t_0<t<3.25\times10^4t_0$ (stage~2) when the interface between the hot inner RIAF and the cool outer region settles around this radius.
The oscillation period is $2\pi/\omega_0=1.0\times10^3t_0$, which is close to the Keplerian rotation period at $r=r_{\mathrm c0}\sim23r_{\mathrm s}$.
It should be noted that azimuthally averaged radial velocity is positive around this radius.
This indicates that the infalling cool blobs are overreflected as shown in the bottom panels of Figure~\ref{fig:terphi}.
}

\textbf{
Another oscillation appears around $r\sim40r_{\mathrm s}$ where the cooling instability grows.
This oscillation can be excited by the vertical contraction of the disk because of radiative cooling.
The vertical contraction induces flattening of the disk as shown in Figure~\ref{fig:snapshots}(e).
Subsequently, the disk thickens (see Figure~\ref{fig:snapshots}(f)), and oscillates mainly in the radial direction.
The amplitude of this oscillation grows with time and becomes prominent when $t>3.5\times10^4t_0$.
The period of the large-amplitude oscillation is $\sim2.0\times10^3t_0$.
This period is close to the Keplerian rotation period at $r\sim40r_{\mathrm s}$.
This oscillation is nonaxisymmetric radial pulsation with an azimuthal mode number $m=1$ as shown in Figure~\ref{fig:terphi}.
}

\textbf{
The black curve in Figure~\ref{fig:lum}(b) shows the time variation of the disk bolometric luminosity $L_{\mathrm z}$ measured at $z=30r_{\mathrm s}$, computed by
%and dashed line in Figure~\ref{fig:luminosity} (b) shows the time variation of the mass accretion rate averaged over $42r_{\mathrm s}<r<44r_{\mathrm s}$.
%
\begin{equation}
 L_\mathrm{z} = \int^{2\pi}_0\int^{r_\mathrm{out}}_{r_\mathrm{in}} F_\mathrm{z}\left(z=30r_{\mathrm s}\right) dr\,rd\varphi, 
 \end{equation}
where $r_\mathrm{in}=0$ and $r_\mathrm{out}=200r_\mathrm{s}$.
Between stage~1 and stage~2, the disk luminosity gradually increases up to roughly $0.5$\% of the Eddington luminosity.
Later, it starts to oscillate quasi-periodically, and the amplitude of the oscillation gradually increases.
The short time scale variabilities observed in the rising stage of CLAGN \citep{noda+2016} can be generated by this oscillation.
The oscillation period of the disk luminosity is close to the oscillation period of the large-amplitude radial pulsation.
% rather than the period of the oscillation excited near the interface between the hot and cool disk.
%
The red curve in Figure~\ref{fig:lum}(b) shows the time variation of the radial velocity averaged over the azimuthal direction in $42r_{\mathrm s}<r<44r_{\mathrm s}$.
The time variation of the radial velocity is almost coherent with the luminosity in $3.0\times10^4t_0<t<3.8\times10^4t_0$ but anticorrelates when $t>3.8\times10^4t_0$ because the luminosity peak coincides with time when the cool blob begins to move outward (see Figure~\ref{fig:terphi} at $t=4.58\times10^4t_0$ and $t=4.78\times10^4t_0$).
}

\textbf{
Figure~\ref{fig:stradi} shows the space--time diagram of azimuthally averaged quantities.
Figure~\ref{fig:stradi}(a) shows the time variability of the surface density.
Figure~\ref{fig:stradi}(b) and (c) shows the space--time diagrams of the radiation pressure, and net emissivity $J = \rho\kappa_{\mathrm{ff}}c \left(a_{\mathrm r}T_{\mathrm{gas}}^4 - E_{\mathrm r}\right)$ averaged in $|z|<0.5r_{\mathrm s}$, respectively (see Figure~\ref{fig:temperature}(b) for temperature).
We take the azimuthal average to focus on the radial oscillation.
The physical quantities oscillate almost coherently in $20r_{\mathrm s}<r<40r_{\mathrm s}$. 
%
%Since the $m=1$ nonaxisymmetric mode grows as Figure~\ref{fig:terphi} shows, this period can be the period of the corotation resonance at $r=r_{\mathrm c1}$.
%
Figure~\ref{fig:stradi}(d) and (e) shows the time variation of the azimuthally averaged kinetic energy in radial and vertical motion integrated in $|z|<2r_{\mathrm s}$.
After the cool region is formed, both the radial and vertical kinetic energy increase.
Figure~\ref{fig:stradi}(f) shows the time variation of the vertical velocity in $z-$time plane.
Between $t=3.0\times10^4t_0$ and $t=4.0\times10^4t_0$, the vertical velocity becomes antisymmetric to the mid-plane.
This antisymmetry is produced due to the vertical contraction by the cooling instability.
}

\textbf{
Figure~\ref{fig:psd} shows the power spectral density (PSD) of the azimuthally averaged radial velocity at the mid-plane.
The PSDs are computed by Fourier transformation of the time evolution of the azimuthally averaged radial velocity at each radius.
Panels (a), (b), and (c) show the PSD in the time interval $1.58\times10^4t_0<t<3.0\times10^4t_0,\  3.05\times10^4t_0<t<3.9\times10^4t_0$, and $3.9\times10^4t_0<t<5.0\times10^4t_0$, respectively.
The green solid and dashed curves show the Keplerian rotation frequency $\Omega_{\mathrm K}/(2\pi)$ and $\kappa/(2\pi)$, respectively, where $\kappa$ is the epicyclic frequency.
The white solid curves show $(\Omega_{\mathrm K}\pm\kappa)/2\pi$ and the yellow solid curves show $(1+\sqrt{2})\Omega_{\mathrm K}/(2\pi)$.
The PSD can be explained by applying the theory of disk oscillation \citep[e.g.,][]{kato2016}.
The local dispersion relation for accretion disks is summarized by \citet{kato2001} as 
\begin{equation}
	\{(\omega-m\Omega_{\mathrm K})^2-\kappa^2\}\{(\omega-m\Omega_{\mathrm K})^2-n\Omega_{\mathrm K}^2\}=k_{\mathrm r}^2c_{\mathrm s}^2(\omega-m\Omega_{\mathrm K})^2,
\end{equation}
where $\omega$ is the wave angular frequency, $n$ is the mode number in the vertical direction, $k_{\mathrm r}$ is the radial wave number, and  $c_{\mathrm s}$ is the sound speed.
This dispersion relation is derived by assuming that the disk is isothermal in the vertical direction near the equatorial plane.
When $n=0$, waves can propagate in the region satisfying $|\omega-m\Omega_{\mathrm K}|>\kappa$.
These are p-mode waves.
When $n\neq0$, because $\kappa<\Omega_{\mathrm K}$, waves can propagate in the region satisfying $-\kappa<\omega -m\Omega_{\mathrm K} <\kappa$ (g-mode) or $|\omega-m\Omega_{\mathrm K}|>\sqrt{n}\Omega_{\mathrm K}$ (c-mode when $n=1$ and vertical p-mode otherwise).
}

\textbf{
In the early stage (Figure~\ref{fig:psd}(a)), oscillations with frequency $\nu>1.0\times10^{-3}t_0^{-1}$ are localized in the hot region ($r<23r_{\mathrm s}$), and the oscillation frequency is less than the epicyclic frequency.
Figure~\ref{fig:psd}(b) shows the stage when the disk shrinks in the vertical direction because of cooling.
As Figure~\ref{fig:stradi}(f) shows, the $n=2$ mode, in which $v_{\mathrm z}$ is antisymmetric to the equatorial plane is excited.
The propagation region of the $m=1$ and $n=2$ g-mode wave is $\Omega_{\mathrm K}-\kappa<\omega<\Omega_{\mathrm K}+\kappa$ (white curves in Figure~\ref{fig:psd}).
The low-frequency oscillation with frequency $\nu=4.0\times10^{-4}t_0^{-1}$ is consistent with this theory.
%
%The contraction of the disk due to cooling also induces radial oscillation.
%
%High frequency oscillations with frequency above $\Omega_{\mathrm K}$ are breathing oscillations \citep[e.g.,][]{blaes+fragile2006,mishra+2019} and p-mode waves. 
%
Another oscillations appear along $\omega=\kappa$.
These are oscillations with local epicyclic frequency.
High--frequency oscillations with frequency above $\Omega_{\mathrm K}$ are p-mode waves.
%
%High--frequency oscillations with frequency above $\Omega_{\mathrm K}$ are breathing oscillations \citep{mishra+2019,mishra+2020}
%
%The PSD in Figure~\ref{fig:psd}(b) is consistent with this theory.
%
%Low frequency wave with frequency $\sim4\times10^{-4}t_0^{-1}$ appears in this region.
%
%The frequency is close to the Keprerian rotation frequency at $r\sim43r_{\mathrm s}$.
%
%The radius is the corotation radius of the $m=1$ nonaxisymmetric mode which becomes evident around $t=4.0\times10^4t_0$ (see Figure~\ref{fig:snapxy}).
%
%On the other hand, the high frequency waves with frequency above $(1+\sqrt{2}\Omega_{\mathrm K})$ (yellow curve) are p-mode waves.
Figure~\ref{fig:psd}(c) shows the later stage when larger amplitude radial oscillation dominates.
%
%Still, we can identify low-frequency ($\nu\sim5.0\times10^{-4}t_0^{-1}$) oscillation confined in $\Omega_{\mathrm K}-\kappa<\omega<\Omega_{\mathrm K}+\kappa$.
%
%In Figure~\ref{fig:psd}(b), another oscillation with frequency $\nu\sim1.0\times10^3t_0$ grows around $r=23r_{\mathrm s}$.
%
%Typical frequency of this oscillation equals to the Keplerian frequency at $r\sim43r_{\mathrm s}$.
%
%Figure~\ref{fig:stradi}(a) indicates that this radius coincides with the outer edge of the dense blob formed by the cooling instability.
%
}

\begin{figure*}
\epsscale{1.0}
\plotone{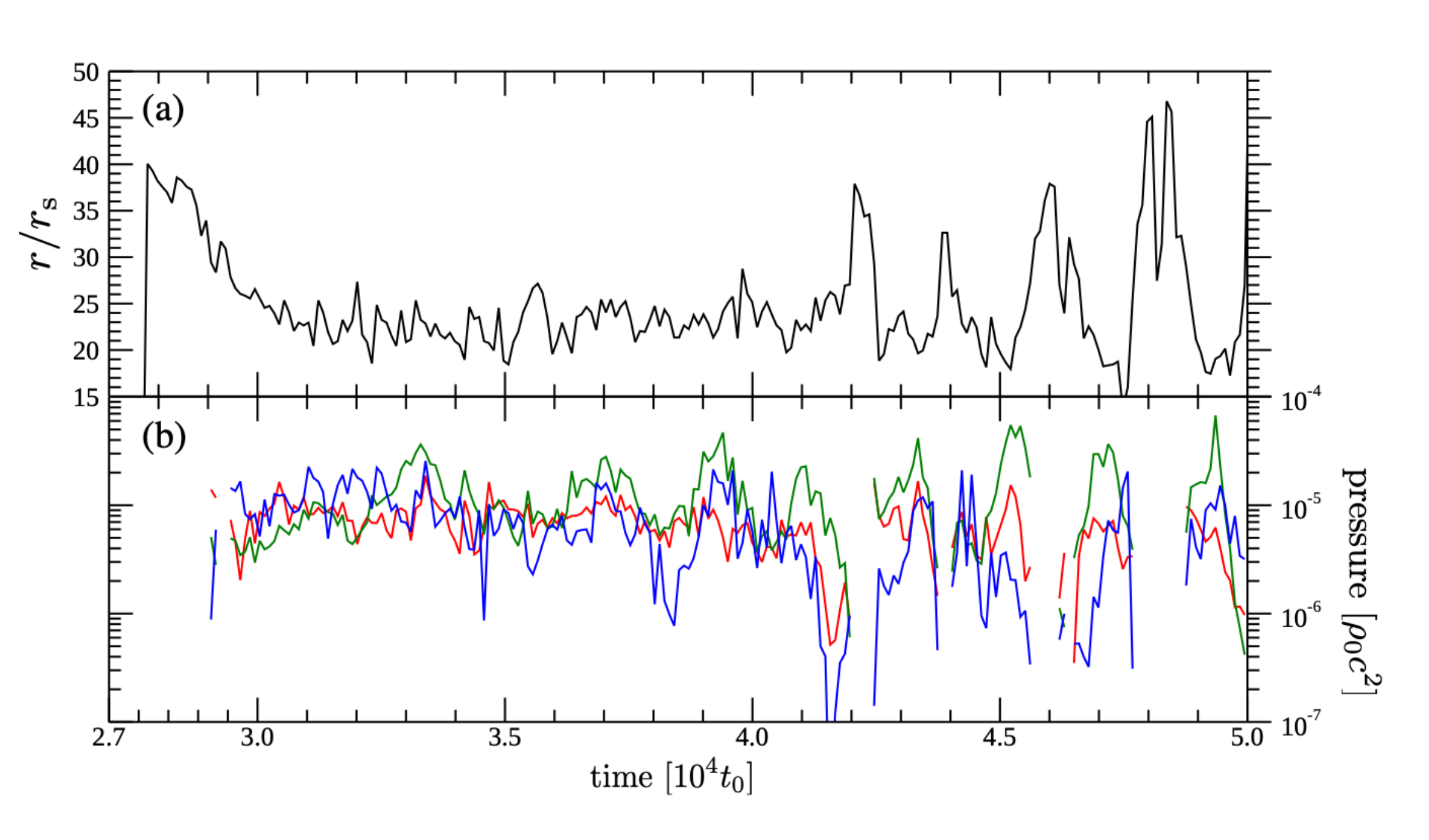}
\caption{Time variation of the (a) innermost radius of the cool region where $T<5\times10^7$ K at the mid-plane, (b) $p_{\mathrm{gas}}$(red), $p_{\mathrm{mag}}$(blue), and $p_{\mathrm{rad}}$(green) averaged in the cool region ($T<5\times10^7$ K) inside $r=30r_{\mathrm s}$ }
\label{fig:cradi}
\end{figure*}
\textbf{
Let us discuss the origin of the large-amplitude radial oscillation.
There can be two mechanisms to amplify the oscillation.
One is the pulsational instability in which the radial oscillation becomes overstable \citep[e.g.,][]{kato1978,blumenthal+1984}.
The other is the corotation instability in which radially propagating p-mode waves are overreflected at the corotation point \citep[e.g.,][]{drury1985,kato1987,tsang+lai2009}.
In Figure~\ref{fig:psd}(b) oscillation with frequency $\nu\sim1.0\times10^{-3}t_0^{-1}$ appears around $r=23r_{\mathrm s}$.
Because $r=r_{\mathrm c0}\sim23r_{\mathrm s}$ is the corotation radius of the $m=1$ wave with angular frequency $\omega_0=\Omega_{\mathrm K}(r_{\mathrm c0})$, a corotation resonance can be responsible for the oscillation at $\omega=\omega_0$ localized around $r=r_{\mathrm c0}$ in Figure~\ref{fig:psd}(b) and (c).
}

\textbf{
It should be noted that g-mode waves are not overreflected at the corotation point \citep{kato2016}.
Because the low-frequency oscillations below $\Omega_{\mathrm K}$ in Figure~\ref{fig:psd} are most likely g-mode waves, they are not amplified by the corotation instability at this frequency.
The low-frequency oscillations with frequency $\nu=5.0\times10^{-4}t_0^{-1}$ in Figure~\ref{fig:psd}(c) correspond to the large-amplitude radial oscillation of the cool region around $r=r_{\mathrm 0}\sim40r_{\mathrm s}$ (see Figure~\ref{fig:terphi}, Figure~\ref{fig:lum} and \ref{fig:stradi}).
%
%The amplitude of the radial oscillation can be amplified by pulsational overstability \citep[e.g.,][]{kato1978,blumenthal+1984}.
%
Figure~\ref{fig:cradi}(a) shows the time variation of the innermost radius of the cool region where $T<5\times10^7$ K at the mid-plane. 
The cool region appears around $t\sim2.7\times10^4t_0$ at $r\sim40r_{\mathrm s}$ and the interface between the hot and cool region moves toward $r\sim20r_{\mathrm s}$.
The radius of the interface is almost constant until $t\sim3.6\times10^4t_0$ but oscillates after this stage.
The amplitude of the displacement of the radius of the cool region increases with time.
Figure~\ref{fig:cradi}(b) shows $p_{\mathrm{gas}}$(red), $p_{\mathrm{mag}}$(blue), and $p_{\mathrm{rad}}$(green) averaged in the cool region ($T<5\times10^7$ K) inside $r=30r_{\mathrm s}$.
The amplitude of the radial pulsation is small until the radiation pressure exceeds the gas pressure and the magnetic pressure. 
Since the growth rate of the radial pulsational instability in radiation pressure dominant disk is larger than that in gas pressure dominant disk \citep[e.g.,][]{blumenthal+1984}, the pulsational instability can be the origin of the amplification of the radial pulsation.
}

\textbf{
For nonaxisymmetric perturbations, \citet{wu+1995a} showed that the growth rate of the nonaxisymmetric radial oscillation increases with the azimuthal mode number in radiation pressure dominant disks.
There is a possibility, however, that strong azimuthal magnetic fields suppress the growth of the nonaxisymmetric radial pulsation. 
The effects of the azimuthal magnetic fields on the radial pulsation has been studied by several authors \citep[e.g.,][]{yang+1995} but their analysis is limited to axisymmetric perturbations.
When the growth of the nonaxisymmetric pulsational instability is suppressed by strong azimuthal magnetic fields, the amplification of the radial pulsation will only be possible when the radiation pressure exceeds the gas pressure and the magnetic pressure.
%
%It will be our future work to prove that azimuthal magnetic field suppresses the growth of the radial pulsational instability in gas + magnetic pressure dominant disks by linear stability analysis.
}

%%%%
\section{Summary and discussion}

%%% Summary
%
In this paper, we presented the results of a global three-dimensional RMHD simulation of AGN accretion flow when $\dot{M} \sim 0.1 \dot{M}_{\rm Edd}$ and our results are as follows;
\begin{itemize}
	\item The soft X-ray emitting, cool ($T\sim10^{7-8}$~K) and dense region forms outside the inner RIAF by radiative cooling.
	\item Radiation pressure becomes dominant in the cool region.
	\item The region has a nonaxisymmetric structure with azimuthal mode number $m=1$.
	\item The region shows quasi-periodic oscillation.
\end{itemize}

\subsection{Effects of Compton scattering}
X-ray observations of Seyfert galaxies indicate that soft X-ray excess component is emitted from the region with electron temperature $T_{\mathrm e}=10^{6-7}$~K and Thomson optical depth $\tau_{\mathrm{es}}\sim10$ \citep[e.g.,][]{done+2012, petrucci+2018}.
These are consistent with the results of our simulations.
In the RMHD simulation we reported in this paper, inverse Compton scattering is not included.
Inverse Compton scattering of soft photons by hot electrons in RIAF produces hard X-rays.
Because the radiation temperature is smaller than the gas temperature except for the mid-plane of the cool region, electrons can be cooled by inverse Compton scattering.
\begin{figure*}
\plotone{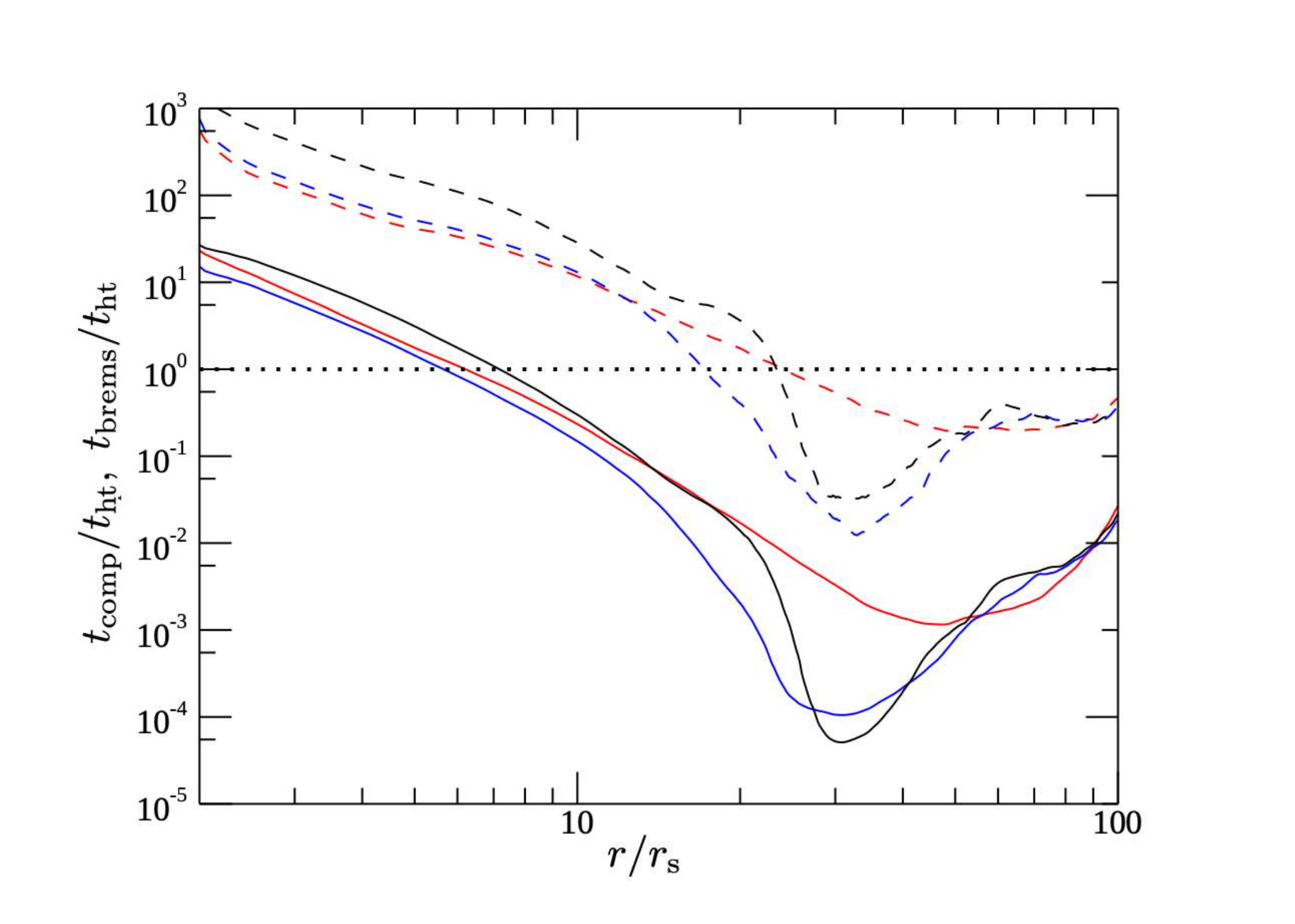}
\caption{The azimuthally averaged radial distribution of the ratio of bremsstrahlung cooling time to viscous heating time (dashed curve) and the Compton cooling time to viscous heating time (solid curve) averaged in $|z|<0.5r_{\mathrm s}$. Red, blue and black curves show stage~1, stage~2 and stage~3, respectively. The dotted horizontal line shows the value of unity.\label{fig:timescale}}
\end{figure*}
\textbf{
Figure~\ref{fig:timescale} shows the radial distribution of the ratio of bremsstrahlung cooling time($t_{\mathrm{brems}}$) to heating time  ($t_{\mathrm{ht}}$) and the Compton cooling time ($t_{\mathrm{comp}}$) to heating time, where $t_{\mathrm{comp}}$ and $t_{\mathrm{ht}}$ are calculated from
\begin{equation}
	t_{\mathrm{comp}} = \frac{p_{\mathrm{gas}}/(\gamma-1)}{\left(\rho\kappa_{\mathrm{es}}cE_{\mathrm r0}\frac{4k_{\mathrm B}(T_{\mathrm e}-T_{\mathrm r})}{m_{\mathrm e}c^2}\right)},
\end{equation}
where $T_{\mathrm r}$ is the radiation temperature and
\begin{equation}
 t_{\mathrm{ht}} = t_{\mathrm{dyn}}/\alpha.
\end{equation}
Here we assume that the electron temperature is $T_{\mathrm e} = \min\left(10^9~\mathrm{K},T\right)$, and $\alpha=0.015$.
%
%In our simulation neglecting the inverse Compton scattering, the cooling time is shorter than the viscous heating time in the region of the dense blob ($r>25r_{\mathrm s}$).
%
}

\textbf{
Figure~\ref{fig:timescale} shows that the time scale of Compton cooling is longer than the viscous heating timescale (= advection time scale) in the hot region where $r<10r_{\mathrm s}$. Therefore, the hot accretion flow will stay hot even when we include the inverse Compton scattering.
This result is consistent with the result of the GRMHD simulation by \citet{takahashi+2016}.
Because the cooling time by inverse Compton scattering becomes shorter than the heating time scale in the region $r>10r_{\mathrm s}$, the region will shrink in the vertical direction faster than our simulation without Compton cooling.
The radius of the interface between the hot inner region and the cool outer region could move inward when inverse Compton scattering is considered.
However, the quasi-steady state of the cool disk will not change even when we consider Compton cooling because the Compton cooling time becomes longer as the electron temperature decreases.
}
%
%When the electrons are cooled down to $10^{6-7}$ K, the Compton cooling rate becomes comparable to the bremsstrahlung cooling and the cooling time becomes longer than the rotation period.
%
We should note that the Compton cooling time shown in Figure~\ref{fig:timescale} is computed using the azimuthally averaged temperature.
As we showed in Figure~\ref{fig:snapxy} (f), the electron temperature can be as low as $10^7$~K.
\textbf{
Because the Compton cooling time is proportional to $1/(T_{\mathrm e}-T_{\mathrm r})$, it becomes $100$ times longer in this region, and becomes comparable to the bremsstrahlung cooling time.
We expect that the gas temperature stays at $10^{6-7}$~K by balancing the dissipative heating and radiative cooling \citep[e.g.,][]{oda+2009,oda+2012}.
The inverse Compton scattering also cools the ambient gases above the disk mid-plane where the Thomson optical depth is larger than 1 (see Figure~\ref{fig:snapshots}).
When the electron temperature is $10^9$ K and $\tau_{\mathrm{es}}=1$, the Compton $y$-parameter is larger than 0.8 so that the inverse Compton scattering is not negligible.
The region will shrink in the vertical direction until the disk is supported by the sum of gas pressure, radiation pressure, and magnetic pressure.
%the radiation pressure and magnetic pressure will increase due to the disk contraction.
%
%When the sum of the radiation pressure and the magnetic pressure becomes comparable to the external gas pressure, the disk will stop shrinking. 
}

%%oscillation
\subsection{Oscillation of the radiation-pressure-dominant region}
\textbf{
The numerical results indicate that a large-amplitude oscillation is exited when the radiation pressure of the cool region exceeds the magnetic pressure and gas pressure.
Such region appears when the luminosity exceeds $0.5$\% of the Eddington luminosity.
The total luminosity of the accretion flow oscillates in coherence with the large-amplitude radial pulsation.
The typical frequency of the large-amplitude pulsation coincides with the Keplerian angular frequency of the dense, cool region.
When the mass of the black hole is $10^7M_{\bigodot}$, the typical period of the oscillation is $2.0\times10^5$ s.
%
%The typical frequency of the large-amplitude pulsation coincides with the Keplerian angular frequency at the outer edge of the dense region formed by cooling.
%%
%In our simulation, this radius locates at $r\sim40r_{\mathrm s}$.
%%
%This radius is close to the initial density maximum so that the radius can be determined by the initial setting.
%%
%%Numerical results indicate that cool region has $m=1$, nonaxisymmetric distribution, and that the cool region oscillates in radial direction between $r=20r_{\mathrm s}$ and $45r_{\mathrm s}$.
%%
%The total luminosity of the accretion flow oscillates in coherence with this large-amplitude radial pulsation.
%%
%The typical period of the oscillation is $2.0\times10^3t_0$ which corresponds to $2.0\times10^5$~s when the mass of the black hole is $10^7M_{\bigodot}$.
%%
}

\textbf{
We should note that the luminosity variation is not because of the rotation of the dense, cool blob because the luminosity is computed by integrating the radiative flux emitted from the disk in $0<r<200r_{\mathrm s}$.
Therefore, the contribution from the nonaxisymmetric distribution is averaged out.
We suggest that the rapid X-ray time variabilities observed during the changing-look phenomena \citep[e.g.,][]{noda+2014} can be because of this oscillation.
Recently, \citet{jin+2020} reported that QPO with period $3.58\times10^3$~s is observed in the narrow line Seyfert 1 galaxy RE J1034+396 whose black hole mass is $\sim2\times10^6M_{\bigodot}$.
This period is shorter than that in our simulations.
We speculate that the oscillation is excited in regions closer to the central black hole.
}

\textbf{
Let us discuss the dependence on the initial density distribution.
If the radial pulsation is amplified when the radiation pressure becomes significant, the oscillation period is determined by the Keplerian rotation period of the radiation-pressure-dominant region.
When the accretion rate is $1-10$\% of the Eddington accretion rate, the radius of the radiation-pressure-dominant region is less than $100r_{\mathrm s}$. 
Therefore, the oscillation frequency is not so different from our simulations.
When the accretion rate is higher, the oscillation period can be longer because the size of the radiation-pressure-dominant region becomes larger.
In such disks, however, thermal instability can collapse the disk to geometrically thin, standard disks as shown by \citet{mishra+2016}.
The quasi-periodic oscillations may occur only when the accretion rate is $1-10$\% of the Eddington accretion rate, so that the radiation-pressure-dominant region is limited to $r<100r_{\mathrm s}$.
}

\textbf{
Considering the above discussions, we need to carry out simulations with different initial density profiles to confirm that the QPOs are excited not at the radius of the initial dense blob but in the region where radiation pressure exceeds the gas pressure and magnetic pressure.
}

%%resolution
\subsection{Final remarks}
In the simulation presented in this paper, we only used $32$ grid points in the azimuthal direction.
\textbf{
\citet{matsumoto+2019} showed that the saturation value of the $\alpha$ is almost independent of the resolution, but the time scale for the saturation becomes shorter in high-resolution simulations.
They also showed that to resolve the growth of MRI, we need at least $128$ grid points in the azimuthal direction.
}
However our essential results do not change even if the azimuthal modes are resolved or if the accretion time scale is changed.
Increasing the azimuthal grid points is difficult because the time step is limited by the CFL condition that becomes short, and computational cost becomes huge.
Nevertheless, we have resolved the nonaxisymmetric structure with mode number $m=1$.
It will be our future work to increase the number of azimuthal grid points.
%

%%outer cold disk
%
In this paper, we have not included the outer cold disk because the accretion time scale from the cold disk is longer than the time scale of our simulation.
To simulate the transition without abruptly including the radiative cooling, it is preferable to simulate the transition by increasing the mass accretion rate from the outer cold disk.
%
%If the outer cold disk is smoothly connected to the inner accretion flow, the oscillation is not excited at this interface, but if the density is discontinuous at the interface of the dense blob formed in the accretion flow and ambient accretion flow, the oscillation is excited.
%
%However, our simulation does not mention the interface between the outer cold disk and inner accretion flow.
%
It will also be our future work to include the outer cold disk.
%
%The cool region correspond to the soft X-ray emitting region. 
%

%%%

\acknowledgments
We thank Mami Machida, Tomohisa Kawashima, Tomoyuki Hanawa, Hideyuki Hotta, and Hirofumi Noda for discussions. This work has been supported in part by
JSPS KAKENHI 16H03954(PI RM), 20H01941(PI RM), 17H01102A(PI KO), 18K03710(PI KO),  JSPS Grant-in-Aid for Young Scientists (17K14260 H.R.T.), 
 and MEXT ``Priority Issue on Post-K Computer (Elucidation of the 
Fundamental Laws and Evolution of the Universe).'' Numerical simulations were performed using 
the XC50 at the Center for Computational Astrophysics, National Astronomical Observatory of Japan. 


%\vspace{5mm}
%\facilities{HST(STIS), Swift(XRT and UVOT), AAVSO, CTIO:1.3m,


%\software{astropy \citep{2013A&A...558A..33A},  
%          Cloudy \citep{2013RMxAA..49..137F}, 
%          SExtractor \citep{1996A&AS..117..393B}
%          }


%% The Equation counter will reset when it encounters the \appendix
%% command and will number appendix Equations (A1), (A2), etc. The

%%%%%

%\appendix
\bibliography{ms}
\bibliographystyle{aasjournal}



\end{document}